\documentclass[hyper]{JHEP3}
\usepackage{amsmath,amssymb,amsfonts,eepic,epsfig,psfrag,mathtools}
\author{V.~A.~Fateev$^{1,2}$ and A.~V.~Litvinov$^{1,3}$ \\
$^1$~Landau Institute for Theoretical Physics, 142432, Russia, Moscow Region, Chernogolovka, acad. Semenov prosp., 1a.\\
$^2$~Laboratoire Charles Coulomb, L2C, UMR 5221, Universit\'e
Montpellier~II, Pl.~E.~Bataillon, 34095 Montpellier, France\\
$^{3}$~Kavli Institute for Theoretical Physics, University of California,\\
 Santa Barbara, CA 93106-4030}
\abstract{In these notes we consider integrable structure of the conformal field theory with the algebra of symmetries $\mathcal{A}=W_{n}\otimes H$, where $W_{n}$ is $W-$algebra  and $H$ is Heisenberg algebra. We found the system of commuting Integrals of Motion with relatively simple properties. In particular, this system has very simple spectrum and the matrix elements of special primary operators between its eigenstates have nice factorized form coinciding exactly with the contribution of the bifundamental multiplet to the Nekrasov partition function for $U(n)$ gauge theories.}
\title{Integrable structure, W-symmetry and AGT relation}
\preprint{NSF-KITP-11-178\\\texttt{\today}}
\dedicated{To the memory of Alyosha Zamolodchikov}
\begin{document}
\section{Introduction}
In the seminal paper \cite{Alday:2009aq} the relation between  $U(2)$ four-dimensional supersymmetric  quiver gauge theories  and two-dimensional conformal field theory with Virasoro symmetry  has been proposed (AGT relation). In \cite{Wyllard:2009hg} the construction of \cite{Alday:2009aq} has been generalized to the relation between  $U(n)$ gauge theories and conformal field theories with $W_{n}$ symmetry. The most non-trivial part of the statement in \cite{Alday:2009aq,Wyllard:2009hg} is the relation between the conformal blocks and the instanton part of the Nekrasov partition function \cite{Nekrasov:2002qd,Nekrasov:2003rj}. This relation has been further checked in some particular cases in \cite{Mironov:2009qt,Mironov:2009by,Alba:2009ya} and was proved in \cite{Mironov:2009qn,Mironov:2010pi,Fateev:2009aw,Hadasz:2010xp} for the special values of the parameters. In our paper \cite{Alba:2010qc} we suggested a proof of AGT relation for linear quiver gauge theories with $U(2)$ gauge groups. The main ingredient of \cite{Alba:2010qc} was the construction of special basis of states in the highest weight representations of the conformal algebra such that the matrix elements of primary operators in this basis have particularly simple factorized form which has natural interpretation in gauge theory. It was shown in \cite{Alba:2010qc} that these states are the eigenvectors for the infinite system of mutually commuting quantities (Integrals of Motion) related to $\text{Benjamin-Ono}_{2}$ integrable hierarchy \cite{Lebedev-Radul,Degasperis,Degasperis2}. This paper is devoted to generalization of our construction to the case of   $W_{n}$ algebras. The logic of the paper is backward to \cite{Alba:2010qc}. We start by defining the set of Integrals of Motion in the universal enveloping of algebra $\mathcal{A}=W_{n}\otimes \mathsf{H}$, where $\mathsf{H}$ is Heisenberg algebra and then find its eigenstates. The matrix elements of appropriate primary fields between these states have again completely factorized form and coincide with the function called $Z_{\textsf{bif}}$.

We note that while in $U(2)$ case the AGT relation gives just another possible algorithm for computation of the conformal blocks comparing to the traditional one \cite{Belavin:1984vu} (see also \cite{Zamolodchikov:1985ie}) for higher rank groups it seems to be the only way which can be accomplished. The standard bootstrap approach to two-dimensional CFT \cite{Belavin:1984vu} requires the knowledge of structure constants of the operator product expansion as well as of the conformal blocks. The situation in $W_{n}$ theory is conceptually more complicated than in the theory with Virasoro symmetry \cite{Bowcock:1993wq}. In particular, the conformal blocks are not fixed by the conformal and $W$-invariances. However, in some particular cases the bootstrap program can be completed similarly to what was done in \cite{Belavin:1984vu}. We consider the $k-$point conformal block on a sphere of special primary fields
\begin{equation}\label{conformal-block}
    \begin{picture}(30,75)(160,10)
    \Thicklines
    \unitlength 2.3pt 
    \put(0,0){\line(1,0){70}}
    \put(72,0){\circle*{.7}}
    \put(74,0){\circle*{.7}}
    \put(76,0){\circle*{.7}}
    \put(78,0){\circle*{.7}}
    \put(80,0){\line(1,0){70}}
    \put(20,0){\line(0,1){25}}
    \put(50,0){\line(0,1){25}}
    \put(100,0){\line(0,1){25}}
    \put(130,0){\line(0,1){25}}
    \put(-7,-1){\mbox{$\alpha_{1}$}}
    \put(16,28){\mbox{$a_{2}\omega_{1}$}}
    \put(46,28){\mbox{$a_{3}\omega_{1}$}}
    \put(94,28){\mbox{$a_{k-2}\omega_{1}$}}
    \put(124,28){\mbox{$a_{k-1}\omega_{1}$}}
    \put(152,-1){\mbox{$\alpha_{k}$}}
    \put(32,3){\mbox{$P_{1}$}}
    \put(58,3){\mbox{$P_{2}$}}
    \put(85,3){\mbox{$P_{k-4}$}}
    \put(110,3){\mbox{$P_{k-3}$}}
    \end{picture}
    \vspace*{1cm}
\end{equation}  
Namely, we took the charges $\alpha_{1}$ and $\alpha_{k}$ to be arbitrary while all the charges $\alpha_{2}\dots\alpha_{k-1}$ to be proportional to the first fundamental weight $\omega_{1}$ of the Lie algebra $sl_{n}$. In all the intermediate channels we took arbitrary fields and parametrized them in terms of momentums $P_{j}$. In this particular case the conformal block \eqref{conformal-block} is completely fixed by the $W$-invariance. At the same time as it was proposed by Wyllard \cite{Wyllard:2009hg} the conformal block in exactly this setup can be related to the instanton part of the Nekrasov partition function for the linear quiver gauge theory with the gauge group 
$$
\underbrace{U(n)\otimes\dots\otimes U(n)}_{k-3}
$$
and with $n-$ fundamental, $n-$ anti-fundamental and $k-4$ bifundamental matter hypermultiplets. The exact relation can be formulated as follows. We use the projective invariance and fix the positions of primary fields in \eqref{conformal-block} as  $z_{1}=0$, $z_{k-1}=1$ and $z_{k}=\infty$. It is also convenient to choose
\begin{equation*}
    z_{i+1}=q_{i}q_{i+1}\dots q_{k-3}\quad\text{for}\quad1\leq i\leq k-3,
\end{equation*}
then the conformal block corresponding to the picture \eqref{conformal-block} is a power series expansion
\begin{equation}\label{conformal-block-explicit}
    \mathbb{F}(q)=
    1+\sum_{\vec{j}}q_{1}^{j_{1}}q_{2}^{j_{2}}\dots q_{k-3}^{j_{k-3}}\,\mathbb{F}_{\vec{j}},
\end{equation}
where sum goes over all set of positive integers $\vec{j}=(j_{1},\dots,j_{k-3})$ and the  coefficients $\mathbb{F}_{\vec{j}}$ are some rational functions of 
$\alpha_{1}$, $\alpha_{k}$, $a_{j}$ and the central charge $c$. The AGT relation claims that the function
\begin{equation}\label{conformal-block-refined}
   \mathbb{Z}(q)\overset{\text{def}}{=}
    \prod_{j=1}^{k-3}\prod_{m=j}^{k-3}(1-q_{j}\dots q_{m})^{a_{j+1}(Q-a_{m+2}/n)}\,\,\mathbb{F}(q)=
    1+\sum_{\vec{j}}q_{1}^{j_{1}}q_{2}^{j_{2}}\dots q_{k-3}^{j_{k-3}}\,\mathbb{Z}_{\vec{j}},
\end{equation}
coincides  the instanton part of the Nekrasov partition function for corresponding gauge theory.
The Nekrasov partition function is known in explicit terms (see \eqref{conformal-block-refined-explicit-coefficient}) and the main ingredient is the contribution of the bifundamental hypermultiplet $Z_{\textsf{bif}}$. In \cite{Alba:2010qc} we constructed a basis of states in highest weight representation of algebra $\mathcal{A}=\textsf{Vir}\otimes\mathsf{H}$ such that the matrix elements  of special primary operators between these states are given exactly by $Z_{\textsf{bif}}$. Here we give the generalization of this construction for $W_{n}$ algebra.

The plan of the paper is the following. In section \ref{Wn} we give short review of $W_{n}$ algebras, then we define the system of commuting Integrals of Motion  in the universal enveloping of algebra $\mathcal{A}=W_{n}\otimes \mathsf{H}$ and then we define primary fields which have factorized matrix elements between eigenstates of this system of IM's. In section \ref{FF} we discuss free field representation of $W_{n}$ algebras and reduce the problem of computation of certain class of the matrix elements to the problem of computing Selberg like contour integrals with insertion of two Jack polynomials. The computation of these integrals is performed in appendix \ref{Selberg-integral}.  In section \ref{Factor} we discuss the factorization properties of the eigenstates which allow to compute the most general matrix elements.  
\section{$W_{3}$ algebra, Integrals of Motion and matrix elements}\label{Wn}
\subsection{$W_{3}$ algebra}
In this paper we consider in details the case of $W_{3}$ algebra. We start by reminding basic definitions. The chiral part of the algebra of symmetries in this case consists of two currents of the spin two $T(x)$ and three $W(z)$
\begin{equation}\label{currents}
    T(z)=\sum_{n=-\infty}^{\infty}\frac{L_n}{z^{n+2}}\qquad
    \text{and}\qquad
   W(z)=\sum_{n=-\infty}^{\infty}\frac{W_n}{z^{n+3}}.
\end{equation}
The Laurent components $L_k$ and $W_k$ form closed $W_3$ algebra with the commutation relations  \cite{Zamolodchikov:1985wn,Fateev:1987vh}
\begin{subequations}\label{W3algebra}
\begin{equation}\label{LunderL}
    \left[L_n,L_m\right]=(n-m)L_{n+m}+\frac{c}{12}(n^3-n)
    \delta_{n,-m},
\end{equation}
\begin{equation}\label{WunderL}
    \left[L_n,W_m\right]=(2n-m)W_{n+m},
\end{equation}
\begin{multline}\label{WunderW}
    \left[W_n,W_m\right]=\frac{c}{3\cdot5!}(n^2-1)(n^2-4)n
    \delta_{n,-m}+\frac{16}{22+5c}(n-m)\Lambda_{n+m}+\\+
    (n-m)\left(\frac{1}{15}(n+m+2)(n+m+3)-\frac{1}{6}(n+2)(m+2)
    \right)
    L_{n+m},
\end{multline}
\end{subequations}
here
\begin{equation*}
    \Lambda_n=\sum_{k=-\infty}^{\infty}:L_kL_{n-k}:+\frac{1}{5}x_n
    L_n,
\end{equation*}
\begin{equation*}
    x_{2l}=(1+l)(1-l)\qquad x_{2l+1}=(2+l)(1-l).
\end{equation*}
Since the Cartan subalgebra of \eqref{W3algebra} is two dimensional it is convenient to introduce vector notations. Namely, let $e_{1}$ and $e_{2}$ be the simple roots of the Lie algebra $sl_{3}$ then it is convenient to parameterize the central charge as 
\begin{equation}
    c=2+12\mathcal{Q}^2=2+24Q^2,\quad\text{where}\quad \mathcal{Q}=Q\rho,\quad Q=b+\frac{1}{b},
\end{equation}
and $\rho$ is the Weyl vector $\rho=e_{1}+e_{2}$. We also parameterize the primary fields $V_{\alpha}$ by the vector parameter $\alpha$.
The operator product expansions of the holomorphic currents \eqref{currents} with the primary fields $V_{\alpha}$ has the form
\begin{equation}\label{OPE}
  \begin{aligned}
    &T(\xi)V_{\alpha}(z)=\frac{\Delta(\alpha)V_{\alpha}(z)}
    {(\xi-z)^2}+
    \frac{\partial V_{\alpha}(z)}{(\xi-z)}+\dots\\
    &W(\xi)V_{\alpha}(z)=\frac{w(\alpha)V_{\alpha}(z)}{(\xi-z)^3}+
    \frac{W_{-1}V_{\alpha}(z)}{(\xi-z)^2}+
    \frac{W_{-2}V_{\alpha}(z)}{(\xi-z)}+\dots
   \end{aligned}
\end{equation}
here\addtocounter{equation}{-1}
\begin{subequations}
\begin{equation}\label{delta}
  \Delta(\alpha)=\frac{(2\mathcal{Q}-\alpha,\alpha)}{2}
\end{equation}
is the conformal dimension and
\begin{equation}\label{omega}
  w(\alpha)=i\sqrt{\frac{48}{22+5c}}\;
  (\alpha-\mathcal{Q},h_1)(\alpha-\mathcal{Q},h_2)(\alpha-\mathcal{Q},h_3)
\end{equation}
\end{subequations}
is the quantum number associated to the $W(z)$ current where $h_{k}$ are wights of the first fundamental representation. 
\subsection{Integrals of Motion}
The computation of the conformal block is equivalent to the problem of computing the matrix elements of primary operator $V_{a\omega_{1}}$ between CFT states. In \cite{Alba:2010qc} we showed that this exercise can be facilitated if one introduces additional bosonic field. In fact in \cite{Alba:2010qc} we considered the conformal field theory with Virasoro symmetry and showed that in the universal enveloping of algebra $\mathcal{A}=\textsf{Vir}\otimes H$  there exists a system of mutually commuting quantities (Integrals of Motion) such that matrix elements of certain primary operators between  its eigenstates have particularly simple factorized form. Now we consider the generalization of our construction to the case of $W_{3}$ algebra. Working with $W-$ algebras it is more convenient to renormalize our additional bosonic field $\phi$. This field does not have zero mode and on a circle can be expanded in the series
\[
\phi(y)=\phi_{+}(y)+\phi_{-}(y)=\sum_{k>0}\frac{ia_{k}}{k}e^{iky}+\sum_{k<0}\frac{ia_{k}}{k}e^{iky}
\]
where operators $a_{k}$ have the commutation relations
\[
[a_{k},a_{l}]=k\delta_{k+l}.
\]
The integral of motion $I$ announced in  \cite{Alba:2010qc} in this case we can be written in the form
\begin{equation}
I_{3}=\sqrt{\frac{8}{3}}\left(6iQ\sqrt{\frac{3}{8}}\sum_{k>0}ka_{-k}a_{k} +\sum_{k\neq0}a_{-k}L_{k}\pm
\frac{\sqrt{4+15Q^{2}}}{4}W_{0}+\frac{1}{3}\sum_{i+j+k=0}a_{i}a_{j}a_{k}\right). \label{I}%
\end{equation}
This integral commutes with the total energy
$$
I_{2}=L_{0}+2\sum_{k=1}a_{-k}a_{k},
$$ 
and also produces an infinite hierarchy of Integrals of Motions. Consider the eigenvectors of this operator at the first level
\begin{equation}
|\Psi_{P}\rangle=(U_{1}W_{-1}+U_{2}L_{-1}+U_{3}a_{-1})|\Theta_{P}\rangle, \label{1}%
\end{equation}
where $|\Theta_{P}\rangle$ is the vacua state such that $L_{n}|\Theta_{P}\rangle=W_{n}|\Theta_{P}\rangle=0$ for $n>0$, 
$L_{0}|\Theta_{P}\rangle=\Delta|\Theta_{P}\rangle$ and $W_{0}|\Theta_{P}\rangle=w_{0}|\Theta_{P}\rangle$.
It is easy to calculate the matrix for the operator \eqref{I}. We introduce the notations
\[
r=1/4\sqrt{4+15Q^{2}},\quad 
\kappa=\frac{2}{4+15Q^{2}},\quad
w_{0}=i\sqrt{\frac{6}{4+15Q^{2}}}\,x_{1}x_{2}x_{3}%
\]
where 
\begin{equation}\label{xj}
 x_{j}=(h_{j},P).
\end{equation}
($h_{i}$ are the weights of fundamental representation of $sl_{3}$ with highest weight $\omega_{1}$). Then taking into account the commutation relations for $W_{3}$ algebra \eqref{W3algebra} our matrix can be written in the form: $M=\frac{2}{\sqrt{3}}M^{\prime}$ where 
$$
M^{\prime}=\begin{pmatrix}
0 & 2r & 0\\
2r\kappa(\Delta-3/4Q^{2}) & 0 & 1\\
3w_{0} & 2\Delta & iQ\frac{3}{2\sqrt{3}}%
\end{pmatrix}
$$
The eigenvalues of this matrix can be easily calculated ($\Delta=Q^{2}-P^{2}/2=Q^{2}+x_{1}x_{2}+x_{1}x_{3}+x_{2}x_{3}$) and are equal
\begin{equation}
\lambda_{i}=i(Q-2x_{i}) \label{ll}%
\end{equation}
They trivially transform under the Weyl reflections $x_{i}\rightarrow x_{j}$ so it is enough to find only one eigenvector. 
For eigenvalue $\lambda_{3}=i(Q-2x_{3})$ it is
\begin{equation}
U_{1}=1,\quad
U_{2}=i\sqrt{\frac{3}{2}}\frac{(Q-2x_{3})}{\sqrt{4+15Q^{2}}},\quad
U_{3}=-\frac{\sqrt{2}(Q-x_{3}+x_{1})(Q-x_{3}+x_{2})}{\sqrt{4+15Q^{2}}}.
\label{Ve}%
\end{equation}
It is easy to see that if we take $P=\omega_{1}y-\mathcal{Q}$ where
$\mathcal{Q}$ $=Q\rho$ ($\rho$ is the Weil vector) we derive that $U_{3}=0$
and our vector $\Psi$ coincides with the null vector
\[
(W_{-1}-\frac{3w_{0}}{_{2\Delta}}L_{-1})|\Theta_{P}\rangle.%
\]
We note that the sign before the operator $W_{0}$ in the integral $I_{3}$ can be chosen arbitrary. The change  of the sign leads to the conjugated representation: $h_{j}\rightarrow-h_{4-j}$. Here we consider the case corresponding to the choice $+$ in \eqref{I}. It is more convenient to renormalize the operators $W_{k}$ in the definition of eigenvectors $\Psi_{P}$. We introduce the operators
\begin{equation}
W_{k}=i\sqrt{\frac{6}{4+15Q^{2}}}\hat{W}_{k}, \label{2}%
\end{equation}
and normalize eigenvectors at the level $k$ by the condition%
\begin{equation}
|\Psi_{P}\rangle=
(\hat{W}_{-1}^{k}+\dots )|\Theta_{P}\rangle, \label{n}%
\end{equation}
then at the first level we will have
\begin{equation}\label{vec1}
|\Psi_{P}\rangle=
\left(  \hat{W}_{-1}+\frac{1}{2}(Q-2x_{3})L_{-1}+\frac{i(Q+x_{13})(Q+x_{23})}{\sqrt{3}}a_{-1}\right)  |\Theta_{P}\rangle%
\end{equation}
where $x_{ij}=x_{i}-x_{j}$. We note that the integral \eqref{I} is  anti-Hermitian under the  conjugation in the algebra $\mathcal{A}=W_{3}\otimes H$ 
\begin{equation}\label{hermitian-conjugation}
W_{-k}^{+}=-W_{k},\qquad\hat{W}_{-k}^{+}=\hat{W}_{k},\qquad L_{-k}^{+}=L_{k},\qquad a_{-k}^{+}=-a_{k},
\end{equation}
and hence the left eigenstates can be defined by the conjugation of the right ones. Moreover, the states corresponding to different eigenvalues are orthogonal to each other.  We can calculate the norm of the vector \eqref{vec1}. It is equal
\begin{equation}
N^{2}=x_{23}x_{13}(Q+x_{23})(Q+x_{13})
\end{equation}
where $x_{ij}=x_{i}-x_{j}$.

At the second level we have nine vectors
\begin{align*}
|\Psi_{P}^{\prime}\rangle  &  =\Bigl(\hat{W}_{-1}^{2}+U_{2}\hat{W}_{-2}+U_{3}L_{-1}\hat{W}%
_{-1}+U_{4}L_{-2}+U_{5}L_{-1}^{2}+U_{6}\hat{W}_{-1}a_{-1}+\\
&  U_{7}L_{-1}a_{-1}+U_{8}a_{-1}^{2}+U_{9}a_{-2}\Bigr)|\Theta_{P}\rangle%
\end{align*}
Now the matrix will be more complicated. Let $M=\frac{2}{\sqrt{3}}M_{1}$ then the matrix $M_{1}$ is
$$%
\begin{pmatrix}
0 & 0 & 2r & 0 & 0 & 0 & 0 & 0 & 0\\
2r\kappa X &0 &  0 & 4r & 2r & 0 & 0 & 0 & 0\\
4r\kappa Y &0 &  0 & 0 & 4r & 1 & 0 & 0 & 0\\
6r\kappa w_{0} &4r\kappa\Delta & 0 & 0 & 0 & 0 & 0 & 0 & 2\\
0 &2r\kappa & 2r\kappa X & 0 & 0 & 0 & 1 & 0 & 0\\
6w_{0} &4 & 2\Delta+2 & 0 & 0 & i3q & 2r & 0 & 0\\
 6\kappa X &0 & 3w_{0} & 3 & 4\Delta+2 & 2r\kappa X & i3q & 2 & 0\\
0 &0 & 0 & 0 & 0 & 6w_{0} & 2\Delta & i6q & 1\\
10\kappa\Delta X &6w_{0} & 9w_{0} &4\Delta+\frac{c}{2}& 6\Delta & 0 & 0 & 1 & i12q
\end{pmatrix}
$$
where $X=(\Delta-\frac{3Q^{2}}{4})$, $Y=(\Delta-\frac{3Q^{2}-2}{4})$ and $q=Q\frac{\sqrt{3}}{2}$.
It is possible to calculate the eigenvalues of this matrix. They have rather simple form linear in $x_{j}$
\[
\lambda_{j}^{(1)}=i(2Q+2x_{j}),\qquad\lambda_{j}^{(2)}=2i(Q+b-2x_{j}),\qquad
\lambda_{j}^{(3)}=2i(Q+1/b-2x_{j})
\]
We can find all the eigenvectors corresponding to these eigenvalues. We give here the eigenvectors for more symmetrical case corresponding to
$\lambda_{j}^{(1)}$. Then the eigenvector for $\lambda_{3}^{(1)}=i(2Q+2x_{3})$
has a form
\begin{align}
U_{2}  &  =-\frac{1}{2}(Q+2x_{3}),U_{3}=(Q+x_{3}),\quad U_{4}=-\frac{(Q+x_{32}%
)(Q+x_{31})}{3},\nonumber\\
U_{5}  &  =\frac{(Q-2x_{2})(Q-2x_{3})}{4},\quad U_{6}=\frac{i(1+Q^{2}+3Qx_{3}%
+x_{12}^{2})}{\sqrt{3}},\nonumber\\
U_{7}  &  =\frac{i(2Q^{3}+5Q^{2}x_{3}+Q(x_{3}^{2}+8x_{2}x_{1}-1)-2x_{3}%
(1+x_{12}^{2}))}{\sqrt{2}},\nonumber\\
U_{8}  &  =U_{4}(Q^{2}-1-4x_{1}^{2}-x_{12}^{2}),\quad U_{9}=iQ\sqrt{3}U_{4}. \label{c1}%
\end{align}
and the norm of this vector is
\[
N^{2}=x_{13}x_{23}(Q+x_{31})(Q+x_{32})(x_{12}+b)(x_{12}-b)(x_{12}%
-1/b)(x_{12}+1/b)
\]
The eigenvectors corresponding to $\lambda_{i}^{(2)}$ and $\lambda_{i}^{(3)}$ are even more simple. For eigenvalue $\lambda_{3}^{(2)}=2i(Q+b-2x_{3})$ they have the following form in the normalization $U_{1}=1$
\begin{align}
U_{2}  &  =-\frac{1}{2}(Q(1+2b^{2})-2b^{2}(x_{3}(2Q+b-2x_{3})-x_{1}%
x_{2})),\quad U_{3}=(Q+b-2x_{3})\nonumber\\
U_{4}  &  =-\frac{b(Q+2b-3x_{3})(Q+x_{23})(Q+x_{13})}{3},\quad U_{5}=\frac
{(Q-2x_{3})(Q+2b-2x_{3})}{4},\nonumber\\
U_{6}  &  =\frac{i2(Q+b+x_{23})(Q+b+x_{13})}{\sqrt{3}},\nonumber\\
U_{7}  &  =i\sqrt{2}(Q-2x_{3})(Q+b+x_{13})(Q+b+x_{23}),\nonumber\\
U_{8}  &  =-\frac{(Q+x_{13})(Q+x_{23})(Q+b+x_{13})(Q+b+x_{23})}{3},\quad 
U_{9}=i\sqrt{3}bU_{8} \label{33}%
\end{align}
The norm of this vector has a form
\begin{align*}
N^{2}  &  =2b(Q-2b)(Q+x_{13})(Q+x_{23})(Q+b+x_{13})(Q+b+x_{23})(b+x_{13})\\
&  \times(b+x_{23})x_{13}x_{23}%
\end{align*}
\subsection{Matrix elements}
For calculation of matrix elements of our vectors we introduce the vertex
operator $V_{a}$
\begin{equation}\label{vertex}
V_{a}=V_{a}^{H}\cdot V_{a}^{W}%
\end{equation}
where
\begin{equation}
V_{a}^{H}=\exp\left(  \phi_{-}(a-3Q)/\sqrt{3}\right)  \exp\left(  \phi
_{+}a/\sqrt{3}\right)  \label{va}%
\end{equation}
and $V_{a}^{W}$ is the primary field of the $W_{3}$ algebra with the charge $\alpha=a\omega_{1}$. This field has the usual commutation relations with the generators of $W$-algebra 
\begin{subequations}\label{A-Vertex-comm-relat}
\begin{equation}\label{W-Vertex-comm-relat}
\begin{aligned}
   &[L_{m},V_{a}^{W}(z)]=z^{m}\Bigl((m+1)\Delta(a\omega_{1})+z\partial_{z}\Bigr)V_{a}^{W}(z),\\
   &[W_{m},V_{a}^{W}(z)]=z^{m}\Bigl(\frac{(m+1)(m+2)}{2}w(a\omega_{1})+z(m+1)W_{-1}+z^{2}W_{-2}\Bigr)V_{a}^{W}(z),
\end{aligned}
\end{equation}
and
\begin{equation}\label{H-Vertex-comm-relat}
\begin{aligned}
\lbrack a_{n}V_{a}^{H}(z)]  &  =-i\sqrt{\frac{1}{3}}a\,V_{a}^{H}(z)z^{n},n<0\\
\lbrack a_{n}V_{a}^{H}(z)]  &  =i\sqrt{\frac{1}{3}}(3Q-a)\,V_{a}^{H}(z)z^{n},n>0
\end{aligned}
\end{equation}
\end{subequations}
The fields $W_{-1}V_{a}^{W}(z)$ and $W_{-2}V_{a}^{W}(z)$ in  \eqref{W-Vertex-comm-relat} are some descendant fields which in general can not be related to the primary field $V_{a}^{W}(z)$ by some differential operator. However, in a special situation which we are considering here, namely, the weight of primary field $V_{a}^{W}(z)$ is proportional to the first fundamental weight, the matrix elements can be computed with the help of \eqref{A-Vertex-comm-relat} and conformal Ward identities (see \cite{Fateev:2007ab} for details). 

Every vector $|\Psi_{\vec{\nu},P}\rangle$ in the highest weight representation of the algebra $\mathcal{A}=W_{3}\otimes H$ can be numerated by three Young diagrams $\vec{\nu}=(\nu_{1},\nu_{2},\nu_{3})$. The vectors at the first level can be denoted as $|\Psi_{(1,0,0),P}\rangle$, $|\Psi_{(1,0,0),P}\rangle$, $|\Psi_{(0,0,1),P}\rangle$, at the second level
$|\Psi_{(2,0,0),P}\rangle$, $|\Psi_{(0,2,0),P}\rangle$, $|\Psi_{(0,0,2),P}\rangle$, $|\Psi_{(\{11\},0,0),P}\rangle$, $|\Psi_{(0,\{11\},0),P}\rangle$, 
$|\Psi_{(0,0,\{11\}),P}\rangle$, $|\Psi_{(1,1,0),P}\rangle$, $|\Psi_{(1,0,1),P}\rangle$, $|\Psi_{(0,1,1),P}\rangle$. For example the vector \eqref{33})is 
$|\Psi_{0,0,\{11\}}\rangle$ and the vector \eqref{c1} is $|\Psi_{1,1,0}\rangle$.

For calculation of the multipoint conformal blocks \eqref{conformal-block} we need the matrix elements
\begin{equation}
F_{\vec{\nu}^{^{\prime}}}^{\vec{\nu}}(a,P,P^{\prime})=\frac{\left\langle
\Psi_{\vec{\nu},P^{\ast}}|V_{a}|\Psi_{\vec{\nu}^{\prime},P^{\prime}%
}\right\rangle }{\left\langle \Psi_{\vec{0},P^{\ast}}|V_{a}|\Psi_{\vec
{0},P^{\prime}}\right\rangle }, \label{me}%
\end{equation}
and also the matrix element
\[
\frac{\left\langle \Psi_{\vec{0},P}|V_{a}|\Psi_{\vec{\nu}^{\prime}P^{\prime}%
}\right\rangle }{\left\langle \Psi_{\vec{0},P}|V_{a}|\Psi_{\vec{0},P^{\prime}%
}\right\rangle },%
\]
here $P^{\ast}$ denotes the usual conjugation: $\Delta(P^{\ast})=\Delta(P)$, $w(P^{\ast})=-w(P)$. The last element differs from $F_{\vec{\nu}'%
}^{\vec{0}}(a,P,P^{\prime})$ by $P\rightarrow-P,$ or $x_{i}\rightarrow-x_{i}$.

It is possible to calculate some matrix elements at the first and second levels. At the first level they are
\begin{align}
F_{100}^{\vec{0}}(a,P,P^{\prime})  &  =-\prod\limits_{j=1}^{3}
(a/3-x_{1}^{\prime}+x_{j})=g_{1}(x,a,x_{1}^{\prime})\nonumber\\
F_{\vec{0}}^{100}(a,P,P^{\prime})  &  =-\prod\limits_{j=1}^{3}
(a/3-(Q-x_{1}+x_{j}^{\prime}))=g_{2}(x_{1},a,x^{\prime})\nonumber\\
F_{100}^{001}(a,P,P^{\prime})  &  =(x_{1}^{\prime}-x_{1}-a/3)(x_{1}^{\prime
}-x_{2}-a/3)(x_{1}^{\prime}-x_{3}+b-a/3)\times\nonumber\\
&  (x_{1}^{\prime}-x_{3}+1/b-a/3)(Q+x_{3}^{\prime}-x_{3}-a/3)(Q+x_{2}^{\prime
}-x_{3}-a/3) \label{l1}%
\end{align}
All other matrix elements can be derived by  simple transformations. At the second
level we calculated elements which appear in the four point conformal block (see below)
\begin{align}
F_{\{11\}00}^{\vec{0}}(a,P,P^{\prime})  &  =\prod\limits_{j=1}^{3}
(a/3-x_{1}^{\prime}+x_{j})(a/3-x_{1}^{\prime}+x_{j}+b)=g_{\scriptscriptstyle{\{1,1\}}}(x,a,x_{1}',b),\nonumber\\
F_{\vec{0}}^{_{00\{11\}}}(a,P,P^{\prime})  &  =\prod\limits_{j=1}^{3}
(a/3-(Q-x_{3}+x_{j}^{\prime}))(a/3-(Q+b-x_{3}+x_{j}^{\prime}))=g^{\scriptscriptstyle{\{1,1\}}}(x_{3},a,x',b),\nonumber\\
F_{110}^{\vec{0}}(a,P,P^{\prime})  &  =\prod\limits_{j=1}^{3}
(a/3-x_{1}^{\prime}+x_{j})(a/3-x_{2}^{\prime}+x_{j})=g_{\scriptscriptstyle{1,1}}(x,a,x_{1}',x_{2}'),\nonumber\\
F_{\vec{0}}^{110}(a,P,P^{\prime})  &  =\prod\limits_{j=1}^{3}
(a/3-(Q-x_{1}+x_{j}^{\prime}))(a/3-(Q-x_{2}+x_{j}^{\prime}))=g^{\scriptscriptstyle{1,1}}(x_{1},x_{2},a,x')\label{l2}%
\end{align}
All other matrix elements having one empty triple of Young diagrams can be derived from these by simple transformations. For example, $\{11\}\rightarrow2$ corresponds to $b\rightarrow1/b$. We note that all the  matrix elements calculated above can be
written as
\begin{equation}
F_{\vec{\nu}^{\prime}}^{\vec{\nu}}(a,P,P^{\prime})=\prod\limits_{i,j=1}^{3}\prod\limits_{s\in\nu_{i}^{\prime}}
(Q-E_{\nu_{i}^{\prime},\nu_{j}}(x_{j}-x_{i}^{\prime}|s)-a/3)\prod\limits_{t\in\nu_{j}}
(E_{\nu_{j},\nu_{i}^{\prime}}(x_{i}^{\prime}-x_{j}|t)-a/3) \label{melm}%
\end{equation}
where
\[
E_{\lambda,\mu}(x|s)=x-bl_{\mu}(s)+(a_{\lambda}(s)+1)/b,
\]
where $a_{\lambda}(s)$ and $l_{\mu}(s)$ are correspondingly the arm length of the square $s$ in the partition $\lambda$ and the leg length of the square $s$ in the partition $\mu$.

It is natural to assume that in $W_{n}-$ case the expression for the matrix elements will have the similar form (here $\vec{\nu}$ is the set of $n$ Young diagrams $\vec{\nu}=(\nu_{1},\dots,\nu_{n})$)
\begin{equation}
F_{\vec{\nu}^{\prime}}^{\vec{\nu}}(a,P,P^{\prime})=\prod\limits_{i,j=1}^{n}\prod\limits_{s\in\nu_{i}^{\prime}}
(Q-E_{\nu_{i}^{\prime},\nu_{j}}(x_{j}-x_{i}^{\prime}|s)-a/n)\prod\limits_{t\in\nu_{j}}
(E_{\nu_{j},\nu_{i}^{\prime}}(x_{i}^{\prime}-x_{j}|t)-a/n). \label{men}%
\end{equation}
We return to this equation later. We note that \eqref{men} coincides exactly with the contribution of the bifundamental multiplet to the Nekrasov partition function \cite{Flume:2002az,Fucito:2004gi,Shadchin:2005cc}
\begin{equation}
  Z_{\textbf{bif}}(\vec{a},\vec{\nu};\vec{b},\vec{\nu'};m),
\end{equation}
where $\vec{a}=(-x_{1},-x_{2},\dots,-x_{n})$, $\vec{b}=(-x'_{1},-x'_{2},\dots,-x'_{n})$, $m=a/n$ and deformation parameters are taken $\epsilon_{1}=b$, $\epsilon_{2}=1/b$.
\subsection{Four-point conformal block}
In order to check our construction we consider four point conformal block of the type \eqref{conformal-block}
\begin{equation}
\left\langle V_{1}(\infty)V_{2}(1)V_{3}(z)V_{4}(0)\right\rangle \label{cb}%
\end{equation}
where fields $V_{1}$ and $V_{4}$ are the primary fields of $W-$ algebra with arbitrary parameters $\alpha_{1},\alpha_{4}$ and fields $V_{2},V_{3}$ have the parameters $\alpha_{i}$ proportional to $\omega_{1}$ or $\omega_{2}$. In this case it is possible to calculate the conformal block
\begin{equation}
\mathrm{F}_{1,2,3,4}(\alpha)=z^{\Delta_{\alpha}-\Delta_{3}-\Delta_{4}}%
(1+f_{1}z+f_{2}z^{2}+\dots ) \label{eb}%
\end{equation}
The first coefficient can be easily calculated and is equal
\begin{align}
f_{1}  &  =\frac{(\Delta_{\alpha}+\Delta_{1}-\Delta_{2})(\Delta_{\alpha
}+\Delta_{3}-\Delta_{4})}{2\Delta_{\alpha}}+\nonumber\\
&  +\left(  \frac{w_{3}-w_{\alpha}}{2}-w_{4}+\frac{3w_{3}}{2\Delta_{3}}%
(\Delta_{\alpha}-\Delta_{4})+\frac{3w_{\alpha}}{2\Delta_{a}}(\Delta_{4}%
-\Delta_{3})\right)  \times\nonumber\\
&  \left(  \frac{-w_{2}-w_{\alpha}}{2}-w_{1}+\frac{3w_{2}}{2\Delta_{2}}%
(\Delta_{1}-\Delta_{a})+\frac{3w_{\alpha}}{2\Delta_{\alpha}}(\Delta_{2}%
-\Delta_{1})\right)  \times\nonumber\\
&  \left(  \Delta_{\alpha}\left[  \frac{4}{4+15Q^{2}}(\Delta_{a}+\frac{1}%
{5})-\frac{1}{5}\right]  -\frac{9w_{\alpha}^{2}}{2\Delta_{\alpha}}\right)
^{-1} \label{f}%
\end{align}

We can calculate independently this function using our combinatorial expansion. In these notes we consider the case when  $\alpha_{2}=a_{2}\omega_{1}$ and $\alpha_{3}=a_{3}\omega_{1}$. In this case we should have in mind that (due to the free field exponents in the vertex operator \eqref{vertex}) it gives us the expansion for the function $\mathrm{F}^{\prime}=(1-z)^{a_{2}(Q-a_{3}/3)}\mathrm{F}$.  For example, for the first coefficient of function $\mathrm{F}_{1,2,3,4}(\alpha)$ we have the following expression in terms of functions $g_{i}$ (\ref{l1})
\begin{equation}
f_{1}=\sum_{k=1}^{3}\frac{g_{1}(-x,a_{2},x_{k}^{\prime})g_{2}(x_{k}^{\prime
},a_{3},y)}{N_{k}^{2}}+a_{2}(Q-a_{3}/3) \label{f1}%
\end{equation}
where $x=\alpha_{1}-\mathcal{Q}$, $y=\alpha_{4}-\mathcal{Q}$, $x_{k}^{\prime}=(h_{k},\alpha-\mathcal{Q})$, 
$$
N_{1}^{2}=x_{12}^{^{\prime}}x_{13}^{^{\prime}}(Q-x_{12}^{^{\prime}})(Q-x_{13}^{^{\prime}})
$$ 
and all other $N_{k}$ can be derived by the cyclic permutations $1\rightarrow2\rightarrow3\rightarrow1\dots$.  It can be checked that expression (\ref{f}) coincides with (\ref{f1}). Similarly, we can obtain
\begin{multline}
    f_{2}=\sum_{i<j}\frac{g_{\scriptscriptstyle{1,1}}(-x,a,x_{i}',x_{j}')g^{\scriptscriptstyle{1,1}}(x_{4-i}',x_{4-j}',a,y)}{N_{i,j}^{2}}+\\+
    \sum_{k=1}^{3}\left(\frac{g_{\scriptscriptstyle{\{1,1\}}}(-x,a,x_{k}',b)g^{\scriptscriptstyle{\{1,1\}}}(x_{4-k},a,y,b)}{(N_{k}(b))^{2}}+
    \frac{g_{\scriptscriptstyle{\{1,1\}}}(-x,a,x_{k}',b^{-1})g^{\scriptscriptstyle{\{1,1\}}}(x_{4-k}',a,y,b^{-1})}{(N_{k}(b^{-1}))^{2}}\right)+\\+
    a_{2}(Q-a_{3}/3)f_{1}+\frac{a_{2}(Q-a_{3}/3)(a_{2}(Q-a_{3}/3)-1)}{2},
\end{multline}
where (together with cyclic permutations)
\begin{equation}
   \begin{aligned}
      &(N_{1,2})^{2}=x_{13}x_{23}(Q+x_{31})(Q+x_{32})(x_{12}+b)(x_{12}-b)(x_{12}%
-1/b)(x_{12}+1/b),\\
      &
      \begin{multlined}
      (N_{1}(b))^{2}=2b(Q-2b)(Q+x_{13})(Q+x_{23})(Q+b+x_{13})(Q+b+x_{23})\times\\
      \times(b+x_{13})(b+x_{23})x_{13}x_{23},\\
      \end{multlined}
   \end{aligned}
\end{equation}
and show that it coincides with the corresponding  coefficient in \eqref{eb}.
\section{Free field representation of $W_{n}$ algebra and Selberg integrals}\label{FF}
\subsection{Bosonization of $W_{n}$ algebra}
We consider here the free field representation of $W_{n}-$ algebra \cite{Fateev:1987vh,Fateev:1987zh} mainly studying this representation for $W_{3}$ current which was denoted as $W$. This current plays the important role because it generates all other currents in the OPE of this current with itself and with new currents. We define the free massless chiral $n-1$ component field $\varphi(z)=\left(
\varphi_{1}(z),\dots ,\varphi_{n-1}(z)\right)  $ normalized by
the condition
\begin{equation}
\,\left\langle \varphi_{i}(z)\varphi_{j}(z^{\prime})\right\rangle
=-\log(z-z^{\prime})\delta_{ij}. \label{n0}%
\end{equation}
and introduce $n-1$ holomorphic currents $\mathrm{W}_{k}(z)$ via the Miura transformation
\begin{equation}
\prod\limits_{i=0}^{n}
:(Q\partial+h_{n-i}\partial\varphi):=\sum_{k=0}^{n}\mathrm{W}_{n-k}(z)\left(Q\partial\right)  ^{k} \label{m}
\end{equation}
where symbol $::$ denotes the Wick ordering and vectors $h_{k}$ are the weights of the first fundamental representation of the Lie algebra
$sl(n)$ with the highest weight $\omega_{1}$
\begin{equation}
h_{k}=\omega_{1}-e_{1}-\dots -e_{k-1},\quad h_{i}h_{j}=1-\frac{1}{n}\delta_{ij}.
\label{hi}
\end{equation}
In particular, it follows from eq (\ref{m}) that $\mathrm{W}_{0}(z)=1,$
$\mathrm{W}_{1}(z)=0$ and the current
\begin{equation}
\mathrm{W}_{2}(z)=T(z)=-\frac{1}{2}(\partial\varphi)^{2}+(\mathcal{Q},\partial^{2}\varphi)=-\frac{1}{2}(\partial\varphi)^{2}+
Q(\rho,\mathcal{\partial}^{2}\varphi)\label{v}%
\end{equation}
where $\rho$ is the Weil vector, form the Virasoro algebra with the central charge
\[
c=n-1+12\mathcal{Q}^{2}=(n-1)\left(  1+n(n+1)Q^{2}\right)
\]
The fields $\mathrm{W}_{k}(z)$ are not conformal primary fields, however their OPEs calculated explicitly in \cite{Lukyanov-1988} contain only bilinear combinations of $\mathrm{W}$- currents. It is easy to check that the field
\begin{equation}
\mathrm{\hat{W}}_{3}(z)=\mathrm{W}_{3}(z)-\frac{(n-2)Q}{2}\partial T(z)
\label{pr}
\end{equation}
will be Virasoro primary field with respect to the stress energy tensor (\ref{v}). The transformation of other fields $\mathrm{W}_{k}(z)$ to the basis of
conformal primaries $\widetilde{\mathrm{W}}_{k}(z)$ is more complicated (see \cite{Watts:1989bn}). The transformation $\mathrm{W}_{k}(z)\rightarrow\widetilde{\mathrm{W}}_{k}(z)$ exists and is invertible but we will not need it here. As all the generators of $W-$algebra depend only on fields 
$(h_{k},\partial\varphi)$ and their derivatives it is convenient to introduce the notation
\begin{equation}
\mathrm{v=}\partial\varphi(z),\quad 
u_{k}(z)=(h_{k},\partial\varphi(z))=(h_{k},\mathrm{v}) \label{u}%
\end{equation}
Sometimes it is convenient to choose the currents $\mathrm{\hat{W}}_{k}(z)$ which transform as 
$\mathrm{\hat{W}}_{k}(z)\rightarrow(-)^{k}\mathrm{\hat{W}}_{k}(z)$
under the natural conjugation $e_{i}\rightarrow e_{n-i}$, $\omega_{i}\rightarrow\omega_{n-i}$, $h_{i}\rightarrow-h_{n+1-i}$. For this we can
define the fields $\mathrm{W}_{k}^{\ast}(z)$ which can be derived from the fields $\mathrm{W}_{k}(z)$ if in their expression in terms of $\partial^{i}u_{k}$ we
make a substitution $u_{k}\rightarrow-u_{n+1-k}$. Then the fields 
$\mathrm{\hat{W}}_{k}(z)=\frac{1}{2}(\mathrm{W}_{k}(z)+(-1)^{k}\mathrm{W}_{k}^{\ast})$ will have the proper transformation under the conjugation. It is easy to check that the field $\mathrm{\hat{W}}_{3}(z)$ defined in such a  way coincides with the primary field defined by eq \eqref{pr}.

The explicit expression for $\mathrm{\hat{W}}_{3}(z)$ in the case $n=3$ has a form
\begin{equation}
\mathrm{\hat{W}}_{3}(z)=:u_{1}u_{2}u_{3}+\frac{Q}{2}\left(  u_{3}\partial
u_{1}-u_{1}\partial u_{3}+\partial((u_{1}-u_{3}\right)  u_{2})-\partial
^{2}u_{2}):. \label{3}%
\end{equation}
Later we will not use the symbol $::$ having in mind that all operators are Wick ordered. It is easy to see from eqs \eqref{m} and \eqref{pr} that  the $\mathrm{\hat{W}}_{3}(z)$ current in  general case can be written as (up to total derivatives)
\begin{equation}
\mathrm{\hat{W}}_{3}(z)=\frac{1}{3}\sum_{i=1}^{n}u_{i}^{3}+Q(u_{n}(\rho_{n-1},\partial\mathrm{v})+u_{n-1}(\rho_{n-2},\partial\mathrm{v})+\dots +u_{3}\partial
u_{1})+\partial(\dots ) \label{der}%
\end{equation}
where $\rho_{k}=(k-1)h_{1}+(k-2)h_{2}+\dots +h_{k-1}$. In particular $\rho_{n}=\rho$, $\rho_{2}=h_{1}$. In eq \eqref{der} we also used the identity
\begin{equation}
\sum_{i>j>k}^{n}u_{i}u_{j}u_{k}=\frac{1}{3}\sum_{i=1}^{n}u_{i}^{3}. \label{in}%
\end{equation}
The usual normalization for $W_{3}-$current $W_{3}(z)=d_{n}^{(3)}%
\mathrm{\hat{W}}_{3}(z)$ is defined by the condition
\begin{equation}
W_{3}(z)W_{3}(z^{\prime})=\frac{c}{3}\frac{1}{(z-z^{\prime})^{6}}+\frac
{2}{(z-z^{\prime})^{4}}T(z^{\prime})+O\left(  \frac{1}{(z-z^{\prime})^{3}%
}\right)  , \label{norm}%
\end{equation}
which defines the coefficient $d_{n}^{(3)}=i\sqrt{\frac{2n}{(n-2)(4+n(n+2)Q^{2})}}$
\begin{equation}
W_{3}(z)=i\sqrt{\frac{2n}{(n-2)(4+n(n+2)Q^{2})}}\mathrm{\hat{W}}_{3}(z).
\label{3n}%
\end{equation}
The important role in the representations of $W-$ algebras play the
exponential fields
\[
\mathrm{V}_{\alpha}(z)=e^{(\alpha,\varphi(z))}.
\]
They are the primary fields of $W-$ algebra i.e.
\[
\mathrm{\hat{W}}_{k}(z)\mathrm{V}_{\alpha}(z^{\prime})=
\frac{\mathrm{w}_{k}(\alpha)}{(z-z^{\prime})^{k}}\mathrm{V}_{\alpha}(z^{\prime})+O\left(\frac{1}{(z-z^{\prime})^{k-1}}\right).
\]
The set of Weyl invariant functions of $P=\mathcal{Q}-\alpha$,  $\mathrm{w}_{k}(\alpha)$ for $k=2,\dots ,n$ characterize the highest weight representations of $W-$ algebras. In particular
\[
\mathrm{w}_{2}(\alpha)=\Delta(\alpha)=\frac{1}{2}(\alpha,\mathcal{Q}%
-\alpha)=\frac{1}{2}(\mathcal{Q}^{2}-\sum_{i}^{n}x_{i}^{2}),\qquad
\mathrm{w}_{3}(\alpha)=\frac{1}{3}\sum_{i}^{n}x_{i}^{3},
\]
where $x_{j}=(h_{j},\alpha-\mathcal{Q})=(h_{j},P)$. Function $\mathrm{w}_{3}(a)$ is related to the function $w_{3}(\alpha)$ defined by \eqref{omega} as
\begin{equation*}
w_{3}(a\omega_{1})=d_{n}^{(3)}\mathrm{w}_{3}(a).
\end{equation*}

All elements of the enveloping of $W-$ algebra commute with the screening charges
\begin{equation}
S_{i}=\int_{C}dz\mathrm{V}_{e_{i}b}(z),\quad\widetilde{S}_{i}=\int
_{C}dz\mathrm{V}_{e_{i/b}}(z) \label{scr}%
\end{equation}
where $C$ is a closed contour. This property permits to calculate some
correlation functions in $W-$ invariant CFT \cite{Fateev:2008bm,Fateev:2005gs,Fateev:2007ab}.

In Hamiltonian approach it is useful to consider the theory on an infinite cylinder of circumference $2\pi$ i.e. to make the conformal transformation $z=e^{-iy}$ where $y=y_{1}+iy_{2}$. The coordinate $y_{2}$ along the cylinder can be interpreted as an imaginary time and $y_{1}\sim y_{1}+2\pi$ as a space coordinate. The fields $T(z)$ and Virasoro primary fields $\mathrm{\hat{W}}_{3}(z)$ under this change of variables transform as
\begin{equation}
T(y)=\frac{c}{24}-z^{2}(y)T(z),\quad\mathrm{\hat{W}}_{3}(y)=(-iz(y))^{3}%
\mathrm{\hat{W}}_{3}(z) \label{ct}%
\end{equation}
The fields $\mathrm{\hat{W}}_{k}(z)$ for $k>3$ defined above in general are not Virasoro primary fields and their transformation has more complicated
form. In principle, we can choose the basis in the $W-$ algebra of conformal primary fields $\widetilde{\mathrm{W}}_{k}(z)$ which transform in a standard way under the conformal transformations. Here we will be mainly interested in fields $T(z)$ and $\mathrm{\hat{W}}_{3}(x)$. We note that after this transformation the fields $\mathrm{\hat{W}}_{k}(y)$ with $k>3$ will still form a $W-$ algebra, but their representation in terms of fields $\mathrm{v}$, contrary to the primary fields, will be modified. We consider the time slice with fixed $y_{2}$. Then all the fields $\mathrm{\hat{W}}_{k}(y)$ can be expanded in Fourier series
\[
\mathrm{\hat{W}}_{k}(y)=\sum_{m=-\infty}^{\infty}\mathrm{\hat{W}}_{k}%
^{(m)}e^{imy}%
\]

The space of states of $W-$ algebra $\mathcal{A}$ is classified by its highest
weight representations.
\[
\mathcal{A=\oplus}_{P}\mathcal{A}_{P}%
\]
Each space $\mathcal{A}_{P}$ contains a primary state $|\Theta_{P}\rangle$ which is annihilated by all positive Fourier modes of operators $\mathrm{\hat{W}}_{k}$ and satisfies
\begin{align*}
\mathrm{\hat{W}}_{k}^{(m)}|\Theta_{P}\rangle  &  =0,\quad m>0,\\
\mathrm{\hat{W}}_{k}^{(0)}|\Theta_{P}\rangle  &  =\mathbf{w}_{k}(P)|\Theta_{P}\rangle
\end{align*}
The free massless field $\mathrm{v}(y)$ and fields $u_{i}(y)$ can be expanded
in Fourier series at the space $\mathcal{A}_{P}$ as
\begin{equation}
\mathrm{v}(y)=i\mathrm{P}-\sum_{m\neq0}\mathrm{b}^{(m)}e^{imy},\quad
u_{i}=ix_{i}-\sum_{m\neq0}c_{i}^{(m)}e^{imy} \label{fff}%
\end{equation}
where the components of the vector $\mathrm{b}^{(m)}=(\mathrm{b}_{1}^{(m)},\dots ,\mathrm{b}_{n-1}^{(m)})$ 
and the operators $c_{i}^{(m)}=(h_{i},\mathrm{b}^{(m)})$ have the commutation relations
\begin{equation}
\lbrack\mathrm{b}_{r}^{(m)},\mathrm{b}_{q}^{(l)}]=m\delta_{m+l,0}\delta
_{r,q},\quad\lbrack c_{i}^{(m)},c_{j}^{(l)}]=m\delta_{m+l,0}(\delta
_{i,,j}-\frac{1}{n}). \label{comr}
\end{equation}
\subsection{Integrals of Motion and Jack polynomials}
We consider now the expression for the integral \eqref{I} and its generalization for the case  of arbitrary $n$ in terms of the fields $u_{k}$
(operators $c_{i}^{(m)}$) and $\phi^{\prime}$ where field $\phi^{\prime}$, has no zero mode and possesses the expansion
\begin{equation}
\phi^{^{\prime}}(y)=-\sum_{k=-\infty}^{\infty}a_{k}e^{iky}. \label{fi}%
\end{equation}
For these purposes we need the expression for the generators $L_{k}$ of the Virasoro algebra and operator 
$W_{3}^{(0)}=id_{n}\frac{1}{2\pi}\int_{0}^{2\pi}\mathrm{\hat{W}}_{3}(y_{1})dy_{1}$. The explicit expression for the
generators $L_{k}$
\[
T(y)=\frac{c}{24}-\sum_{m=-\infty}^{\infty}L_{k}e^{iky},%
\]
where
\begin{equation}
\begin{gathered}
\begin{multlined}
L_{k\neq0}=\frac{1}{2\pi}\int\limits_{0}^{2\pi}\left(  \frac{1}{2}\sum_{i=1}^{n}u_{i}^{2}-Q(n-i)u_{i}^{\prime}\right)
e^{-iky}=\\=
\frac{1}{2}\sum_{j=1}^{n}\sum_{m\neq k,0}c_{j}^{(m)}c_{j}^{(k-m)}+i\sum_{j=1}^{n}(kQ(n-j)-x_{j})c_{j}^{(k)},\\
\end{multlined}\\
L_{0}=\sum_{j=1}^{n}\left(-\frac{1}{2}x_{j}^{2}+\sum_{m>0}^{\infty}c_{j}^{(m)}c_{j}^{(m)}\right).  \label{lboz}%
\end{gathered}
\end{equation}
To write the explicit expression for $\mathrm{\hat{W}}_{3}^{(0)}$ (or
$W_{3}^{(0)}$) operator it is convenient to introduce the fields
$\widetilde{u}_{k},$ $\widetilde{\mathrm{v}},$ which do not have the zero modes
\[
u_{k}(y)=ix_{k}+\widetilde{u}_{k}(y),\quad\mathrm{v}(y)=i\mathrm{\hat{P}%
}+\widetilde{\mathrm{v}}(y).
\]
Sometimes we will also use the notations 
$$
\widetilde{u}_{k}^{(+)}(y),\quad
\phi_{+}^{\prime}(y),\quad
(\widetilde{u}_{k}^{(-)}(y),\phi_{-}^{\prime}(y)),
$$ 
where
$u_{k}(y)=$ $\widetilde{u}_{k}^{(+)}(y)+\widetilde{u}_{k}^{(-)}(y)$ and fields $\widetilde{u}_{k}^{(+)}(y)$, 
$\phi_{+}^{\prime}(y)$ and $(\widetilde{u}_{k}^{(-)}(y),\phi_{-}^{\prime}(y))$ contain only positive (negative) Fourier components of the fields $u_{k}(y)$ and $\phi^{\prime}(y)$. For example, the first and second terms in the integral $I_{3}^{(n)}$ (\ref{In}) can be written as
\begin{align}
\frac{n\sqrt{n}Q}{2}i\sum_{k=1}^{\infty}ka_{-k}a_{k}  &  =\frac{n\sqrt{n}Q}%
{2}\frac{1}{2\pi}\int\limits_{0}^{2\pi}
dy_{1}\phi_{-}^{\prime}(y_{1})\phi_{+}^{\prime\prime}(y_{1}).\nonumber\\
\sum_{k=-\infty}^{\infty}ka_{k}L_{-k}  &  =-\frac{1}{2\pi}\int\limits_{0}^{2\pi}
dy_{1}\phi^{\prime}(y_{1})T(y_{1})=\nonumber\\
&  -\frac{1}{2\pi}\int\limits_{0}^{2\pi}dy_{1}\phi^{\prime}\sum_{k=1}^{n}\frac{1}{2}\left(  \widetilde{u}_{k}%
^{2}+2(ix_{k}-Q(n-k))\widetilde{u}_{k}^{\prime}\right)  \label{tr}%
\end{align}
The operator $\mathrm{\hat{W}}_{3}^{(0)}$ in these notation has a form \eqref{der}, \eqref{ct}
\begin{equation}
i\mathrm{\hat{W}}_{3}^{(0)}=\int\limits_{0}^{2\pi}dy_{1}\left(  \frac{1}{3}\sum_{k=1}^{n}\left(  \left(  ix_{i}^{3}%
-3ix_{i}\widetilde{u}_{k}^{2}-\widetilde{u}_{k}^{3}\right)  -Q\widetilde
{u}_{n+1-k}\rho_{n-k}\widetilde{\mathrm{v}}^{\prime}(y_{1})\right)  \right)
\label{wr}
\end{equation}
where 
$$
(\rho_{n-k},\widetilde{\mathrm{v}}^{\prime}(y_{1}))=(n-k-1)\widetilde{u}_{1}^{\prime}+(n-k-2)\widetilde{u}_{2}^{\prime}+\dots +\widetilde{u}%
_{n-k-1}^{\prime},\quad
(\rho_{1},\widetilde{\mathrm{v}}^{\prime}(y_{1}))=0.
$$ 
We should have in mind that all our expressions are Wick ordered i.e. the fields
$\widetilde{u}_{k}^{(+)}(y)$ are always placed at the right from the fields
$\widetilde{u}_{k}^{(-)}(y)$. In terms of fields $\widetilde{u}_{k},\phi^{\prime}$ the generalization of integral $I_{3}$ \eqref{I} the operator
$I_{3}^{(n)}=\frac{2}{\sqrt{n}}\mathrm{I}_{3}^{(n)}$ can be written as
\begin{align}
\mathrm{I}_{3}^{(n)}  &  =\frac{\sqrt{4+n(n+2)Q^{2}}\sqrt{n-2}}{2^{3/2}}W_{3}^{(0)}+
\frac{1}{2\pi}\int\limits_{0}^{2\pi}dy_{1}\left(  \frac{n\sqrt{n}Q}{2}\phi_{-}^{\prime}\phi_{+}^{\prime\prime
}-\phi^{\prime}T-\frac{\phi^{\prime3}}{6}\right) =\nonumber\\
&  =i\frac{\sqrt{n}}{2}\mathrm{\hat{W}}_{3}^{(0)}+
\frac{1}{2\pi}\int\limits_{0}^{2\pi}dy_{1}\left(  \frac{n\sqrt{n}Q}{2}\phi_{-}^{\prime}\phi_{+}^{\prime\prime
}-\phi^{\prime}T-\frac{\phi^{\prime3}}{6}\right).\label{In}%
\end{align}
To study the integral $\mathrm{I}_{3}^{(n)}$ we introduce the fields
\[
\Phi_{k}(y)=\Phi_{k}^{(+)}(y)+\Phi_{k}^{(-)}(y)=\frac{\phi^{\prime}+\sqrt{n}\widetilde{u}_{k}}{\sqrt{n}}=-\sum_{m\neq0}A^{(m)}_{k}e^{imy}%
\]
where operators $A_{k}^{(m)}=\frac{a_{m}+\sqrt{n}c_{k}^{(m)}}{\sqrt{n}}$ have
the commutation relations
\[
\lbrack A_{i}^{(m)},A_{j}^{(k)}]=m\delta_{0}^{m+k}\delta_{j}^{i}%
\]
Our operator $\mathrm{I}_{3}^{(n)}$ acts in the Fock space $\mathcal{F}_{P}$ generated by these operators from the vacuum state $|P\rangle$, which is
annihilated by all the operators $\Phi_{i}^{(+)}$  and satisfies the equation $\mathrm{P}|P\rangle=P|P\rangle$. Now we show that the integral
$\mathrm{I}_{3}^{(n)}$ has in this space the invariant subspace $F_{n},_{M}$ generated by the operators $A_{n}^{(m)}$ with negative $m$ of the form
\[
F_{n},_{M}=\sum_{k}\sum_{m_{1}+\dots +m_{k}=M}C^{(m_{1},\dots  ,m_{k})}A_{n}%
^{(-m_{1})}\dots  A_{n}^{(-m_{k})}|P\rangle
\]
which is annihilated by all operators $\Phi_{i}^{(-)}$ with $i\neq n$.

We  start with terms which depend on $x_{i}$. We will not take into account
the constant cubic term $\frac{i}{3}\sum_{1}^{n}x_{i}^{3}$. It follows from
the eqs \eqref{tr}, \eqref{wr} and \eqref{In} that these terms are
\begin{align*}
\frac{-i}{2\pi}\int\limits_{0}^{2\pi}
\sum_{k=1}^{n}\frac{x_{k}}{2}\left(  2\phi^{\prime}\widetilde{u}_{k}+\sqrt
{n}\widetilde{u}_{k}^{2}\right)  dy_{1}  &  =\frac{-i}{2\pi}\int\limits_{0}^{2\pi}
\sum_{k=1}^{n}\frac{x_{k}}{2}\left(  \phi^{\prime2}/\sqrt{n}+2\phi^{\prime
}\widetilde{u}_{k}+\sqrt{n}\widetilde{u}_{k}^{2}\right)  dy_{1}=\\
\frac{-i}{2\pi}\int\limits_{0}^{2\pi}
\sum_{k=1}^{n}\frac{x_{k}\sqrt{n}}{2}\Phi_{k}{}^{2}  &  =\frac{-i}{2\pi}\int\limits_{0}^{2\pi}
\sum_{k=1}^{n}x_{k}\sqrt{n}\Phi_{k}{}^{(-)}\Phi_{k}{}^{(+)}.%
\end{align*}
We note that this operator acts on $F_{n},_{M}$ as Casimir with eigenvalue $-ix_{n}\sqrt{n}M$.

It easy to see that the cubic terms in the integral $\mathrm{I}_{3}^{(n)}$ can be written as
\begin{align*}
\frac{-1}{2\pi}\int\limits_{0}^{2\pi}
\frac{dy_{1}}{6}\left(  \phi^{\prime3}+\sum_{k=1}^{n}(3\phi^{\prime}%
\widetilde{u}_{k}^{2}+\sqrt{n}\widetilde{u}_{k}^{3})\right)   &  =\frac
{-1}{2\pi}\int\limits_{0}^{2\pi}
\sum_{k=1}^{n}\frac{dy_{1}\sqrt{n}}{6}\Phi_{k}{}^{3}=\\
&  =\frac{-1}{2\pi}\int\limits_{0}^{2\pi}
\frac{dy_{1}\sqrt{n}}{2}\sum_{k=1}^{n}\left(  \Phi_{k}{}^{(-)}{}^{2}\Phi_{k}%
{}^{(+)}+\Phi_{k}{}^{(-)}{}\Phi_{k}{}^{(+)2}\right)
\end{align*}
We see that the space $F_{n},_{M}$ is invariant under the action of this operators.

To study the quadratic terms it is convenient to introduce the notation
$\Phi_{0}=\Phi_{1}+\dots +\Phi_{n-1}=\frac{(n-1)\phi^{\prime}-\sqrt{n}u_{n}%
}{\sqrt{n}}$. Field $\Phi_{0}$ commutes with $\Phi_{n}$ and $\phi^{\prime
}=\frac{1}{\sqrt{n}}(\Phi_{0}+\Phi_{n})$. The quadratic terms have the form
\begin{equation}
\frac{Q}{2\pi}\int\limits_{0}^{2\pi}dy_{1}\left(  \frac{\sqrt{n}}{2}(\Phi_{0}^{(-)}+\Phi_{n}^{(-)})\partial
(\Phi_{0}^{(+)}+\Phi_{n}^{(+)})+\phi^{\prime}\rho\widetilde{\mathrm{v}%
}^{\prime}-\sum_{1}^{n}\frac{\sqrt{n}}{2}u_{n-k+1}\rho_{(n-k)}\widetilde
{\mathrm{v}}^{\prime}\right)  \label{qad}%
\end{equation}
After simple transformations neglecting the total derivatives we arrive to the
following expression for (\ref{qad})
\begin{equation}
\frac{Q}{2\pi}\int\limits_{0}^{2\pi}dy_{1}\left(  \frac{\sqrt{n}}{2}(\Phi_{0}^{(-)}+\Phi_{n}^{(-)})\partial
(\Phi_{0}^{(+)}+\Phi_{n}^{(+)})+\frac{\sqrt{n}}{2}\Phi_{0}^{\prime}\Phi
_{n}+S_{n-1}\right)  \label{q1}%
\end{equation}
where the term $S_{n-1}$ annihilates the vectors in $F_{n},_{M}$ and has a form
\[
S_{n-1}=\sum_{i>j}^{n-1}\frac{\sqrt{n}}{2}\Phi_{i}\Phi_{j}^{\prime}.
\]
The \textquotedblleft unwanted\textquotedblright\ term in the first term of
integrand in eq (\ref{q1}) is $\Phi_{0}^{(-)}\partial\Phi_{n}^{(+)}$ but it
cancels after integration. The first two terms in the integrand in eq
(\ref{q1}) can be written as
\[
\frac{\sqrt{n}}{2}\left(  \Phi_{n}^{(-)}\partial\Phi_{n}^{(+)}+\Phi_{0}%
^{(-)}\partial\Phi_{0}^{(+)}+\Phi_{n}^{(-)}\partial\Phi_{0}^{(+)}+\Phi
_{0}^{(-)}\partial\Phi_{n}^{(+)}+\Phi_{n}^{(-)}\partial\Phi_{0}^{(+)}%
+\partial\Phi_{0}^{(-)}\Phi_{n}^{(+)}\right)  .
\]
We see that the terms $\Phi_{0}^{(-)}\partial\Phi_{n}^{(+)}+\partial\Phi
_{0}^{(-)}\Phi_{n}^{(+)}$ form the total derivative and cancel after
integration. If we take into account the terms $\Phi_{0}^{(-)}\partial\Phi
_{0}^{(+)}$, $2\Phi_{n}^{(-)}\partial\Phi_{0}^{(+)}$ and the term $S_{n-1}$ the
quadratic terms in the integrand can be written in the form
\[
\frac{\sqrt{n}}{2}\left(  \sum_{k=1}^{n}\Phi_{k}^{(-)}\partial\Phi_{k}%
^{(+)}+2\sum_{i>j}^{n}\Phi_{i}^{(-)}\partial\Phi_{j}^{(+)}\right).
\]
Taking into account all terms the expression for the integral $I_{3}^{(n)}=\frac{2}{\sqrt{n}}\mathrm{I}_{3}^{(n)}$ can be written as
\begin{equation}
I_{3}^{(n)}=\frac{1}{2\pi}\int\limits_{0}^{2\pi}dy_{1}\left(  \sum_{k=1}^{n}(Q\Phi_{k}^{(-)}\partial\Phi_{k}^{(+)}-2ix_{k}%
\Phi_{k}{}^{(-)}\Phi_{k}{}^{(+)}-\frac{1}{3}\Phi_{k}^{3})+2Q\sum_{i>j}^{n}%
\Phi_{i}^{(-)}\partial\Phi_{j}^{(+)}\right)  \label{inf}%
\end{equation}
We note that under Hermitian conjugation, which acts as $i\Phi_{j}\rightarrow i\Phi_{n+1-j}$ the first three terms change the sign and the last term changes also the order in the sum over $i,j$. It means that the left invariant subspace will be the space $F_{1,M}^{\ast}$. 

It is worth to mention that the operator $I_{3}^{(n)}$ possesses nice representation in terms of differential operators. Namely, consider the quotient of the integral $I_{3}^{(n)}$ on invariant subspace $F_{n},_{M}$
\begin{equation}
   \hat{I}_{3}^{(n)}=\frac{1}{2\pi}\int\limits_{0}^{2\pi}dy_{1}\left( Q\Phi_{n}^{(-)}\partial\Phi_{n}^{(+)}-2ix_{n}%
\Phi_{n}{}^{(-)}\Phi_{n}{}^{(+)}-\frac{1}{3}\Phi_{n}^{3}\right),  \label{inf-quation}
\end{equation}
and replace
\begin{equation}
   A_{n}^{-k}=-ibp_{k},\qquad
   A_{n}^{k}=\frac{i}{b}k\frac{\partial}{\partial p_{k}},
\end{equation}
where $p_{k}=p_{k}(z)$ are power-sum symmetric polynomials in infinitely many variables $z_{j}$
\begin{equation*}
   p_{k}(z)=\sum z_{j}^{k}.
\end{equation*}
When acting on a space of symmetric functions the operator $\hat{I}_{n}$ can be rewritten as a differential operator
\begin{equation}
   \hat{I}_{n}=ib^{-1}\left(\mathcal{H}_{2}+\frac{b^{2}}{2}\mathcal{H}_{1}^{2}-bx_{n}\mathcal{H}_{1}\right),
\end{equation}
where $\mathcal{H}_{1}$ and $\mathcal{H}_{2}$ are first two Calogero-Sutherland Hamiltonians
\begin{equation}
  \begin{aligned}
     &\mathcal{H}_{1}=\sum z_{i}\frac{\partial}{\partial z_{i}},\\
     &\mathcal{H}_{2}=\sum \left(z_{i}\frac{\partial}{\partial z_{i}}\right)^{2}+g\sum_{i<j}\frac{z_{i}+z_{j}}{z_{i}-z_{j}}
     \left(z_{i}\frac{\partial}{\partial z_{i}}-z_{j}\frac{\partial}{\partial z_{j}}\right),
  \end{aligned}
\end{equation}
with $g=-b^{2}$. Similarly, we can shown that the operator $I_{3}^{(n)}$ restricted on a left invariant subspace $F_{1,M}^{\ast}$ can also be rewritten as a Calogero-Sutherland Hamiltonian. One can also show that higher integrals of motion are reducible to higher CS Hamiltonians.
All this allows to prove that the right and the left eigenvectors of this operator can be expressed through the Jack polynomials associated with
Young diagrams $\vec{\nu}=(0,\dots  0,\nu_{n})$ and $\mu=(\mu_{1},0,\dots  0)$. Namely, the states $|P^{\prime}\rangle_{\vec{\lambda}}$ and $_{\vec{\mu}}\langle P^{\ast}|$ corresponding to these diagrams can be defined as
\begin{align}
|P^{\prime}\rangle_{_{\vec{\nu}}}  &  =\Omega_{\nu_{n}}(P^{\prime}%
)\mathrm{J}_{_{\nu_{n}}}^{(1/g)}(y)|P^{\prime}%
\rangle\nonumber\\
_{\vec{\mu}}\langle P^{\ast}|  &  =\Omega_{\mu_{1}}(P)\langle P^{\ast
}|\mathrm{J}_{\mu_{1}}^{(1/g)}(x) \label{j}%
\end{align}
where $g=-b^{2}$,
\[
A_{n}^{(-m)}=-ibp_{m}(y),\quad
A_{1}^{(m)}=ibp_{m}(x)
\]
and $\ \mathrm{J}_{\nu}^{(1/g)}(x)$ is the Jack polynomial
associated with the diagram (partition) $\nu$ normalized as
\[
\ \mathrm{J}_{\nu}^{(1/g)}(x)=|\nu|!m_{[1\dots 1]}(x)+\dots 
\]
where $m_{[\nu_{1}\dots ]}$ is the monomial symmetric polynomial. The
factors $\Omega_{\nu_{k}}(P)$ ($k=1,n)$ are defined by
\begin{equation}
\Omega_{\nu_{k}}(P)=(-b)^{|\nu_{k}|}\prod\limits_{l\neq k}\prod\limits_{(i,j)\in\nu_{k}}
(x_{lk}+ib+jb^{-1}) \label{om}%
\end{equation}
where index $i$ runs horizontally and $j$ vertically over diagram $\nu_{k}$
and $x_{lk}=x_{l}-x_{k}$.

Before proceed further we note that if we take parameters $x_{k}$ in eq \eqref{inf} $|x_{k}|\gg |Q|$ we can neglect the first and last quadratic
terms in the integral of motion $I_{3}^{(n)}$. In this limit the eigenvectors of $I_{3}^{(n)}$ will not depend on $b$ and can be written as (\textquotedblleft asymptotic freedom\textquotedblright)
\begin{equation}
|P\rangle_{\vec{\nu}}=\prod\limits_{i=1}^{n}
\mathrm{J}_{\nu_{i}}^{(1)}(A_{i}^{(-1)},A_{i}^{(-2)}\dots )|P\rangle\label{af}%
\end{equation}
This asymptotic gives the indication for the appearance of Young diagram in the classification of eigenstates of $I_{3}^{(n)}$ which can be considered as finite
momentum deformation of the states \eqref{af}. The eigenvalues of operator  $I_{3}^{(n)}$ will have the asymptotics $\mathfrak{Q}_{\vec{v}}^{(3)}=\sum_{j=1}^{n}-2i|\nu_{j}|x_{i}$ which is the sum over all partitions $\nu_{j}$. It is possible calculate the first correction in $Q$ to the spectrum and find that it is also the sum over all partitions. If we take $|P\rangle_{_{\vec{\nu}}}$ in the form (\ref{j}) we will have $\mathfrak{Q}%
_{\vec{v}}^{(3)}=q^{(3)}(\nu_{n},x_{n})$ where
\begin{equation}
q^{(3)}(\nu,x)=i\left(  -2|\nu|x+\frac{1}{b}\sum_{l}\nu_{l}(\nu_{l}+(2l-1)b^{2})\right)  . \label{eig}%
\end{equation}
Our integral \eqref{In} does not depend on $x_{i}$. This dependence appear after bosonization. There exist $n!$ different bosonizations corresponding to different Weyl transformations of the vector $P$. This group acts as the group of permutation $S_{n}$ of the parameters $x_{k}$. It is always possible to choose the transformation of the Weyl group $\hat{s}$ such that the operator $\mathfrak{Q}^{(3)}$ written in terms of operators $A_{k}^{(-m)}(\hat{s}),$ which possess rather complicated expansion in terms of operators $A_{j}$ (all $A_{k}^{(-m)}(\hat{s})$ have canonical commutation relations and hence connected by canonical transformation) will have the invariant subspace $F_{i,M}$ with arbitrary $i$ and eigenstate (\ref{j}) with eigenvalue $q^{(3)}(\nu_{i},x_{i})$. All this gives us the reasons to assume that operator $\mathfrak{Q}^{(3)}$ similar to $sl(2)$ case  \cite{Alba:2010qc} has the spectrum
\begin{equation}
\mathfrak{Q}_{\vec{\nu}}^{(3)}=\sum_{j=1}^{n}q^{(3)}(x_{j},\nu_{j}).
\label{ras}%
\end{equation}
Of course this property is valid for the first integral $I_{2}^{(n)}=L_{0}+\sum_{k>0}a_{-k}a_{k}$. It can be proved that it is valid
for all higher integrals $\mathfrak{Q}^{(s)}$, $s=4,5,\dots $.
\subsection{Free-field representation of the matrix elements and Selberg integrals}
The important property of operators $A_{1}$ and $A_{n}$ is that they commute and have simple commutation relations with vertex operator which for $W_{n}$ case has a form
\begin{equation}
\mathcal{V}(z)=e^{(a-Qn)\phi_{-}(z)/\sqrt{n}}e^{a\phi_{+}(z)/\sqrt{n}}\mathrm{V}_{a\omega_{1}}(z).
\end{equation}
Namely, operators $A_{n}$ commute with $\mathcal{V}$ and $[A_{1}%
^{(m)},\mathcal{V}(1)]=i(a-Q)\mathcal{V}(1)$.

\begin{figure}
\psfrag{a}{$0$}
\psfrag{b}{$1$}
	\centering
	\includegraphics[width=.6\textwidth]{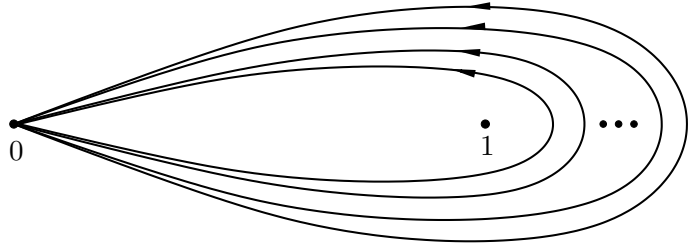}
        \caption{Contours of integration in the integral \protect\eqref{ev}.}
	\label{contours}
\end{figure}

With these properties we can reduce the calculation of function $F_{\scriptscriptstyle{\varnothing}
,\dots  ,\scriptscriptstyle{\varnothing},\nu^{\prime}}^{\nu,\scriptscriptstyle{\varnothing},\dots  ,\scriptscriptstyle{\varnothing}}(a,P,P^{\prime})$ for
special values of parameters $P^{\prime}$ and $P^{\ast}=-P$ to the $sl(n)$
analog of Selberg integral \cite{Tarasov-Varchenko,Warnaar-An}. Namely, if vectors
$P^{\ast}=-P,P^{\prime},$ in eq \eqref{me} satisfy the condition
\begin{equation}
-P+P^{\prime}+\omega_{1}a+\sum_{s=1}^{n-1}bk_{r}e_{r}=0 \label{sc}%
\end{equation}
where $e_{k}$ are the simple roots of $sl(n)$ and $k_{r+1}\leq k_{r}$ are
non-negative integers, the calculation of matrix element $F_{(0,\dots  0,\mu
)}^{(\lambda,0,\dots  ,0)}(a,P,P^{\prime})$ can be reduced to the calculation of
the integral with screening charges (\ref{scr}) $S_{i}$. The operators
$A_{n}^{(-m)}$ commute with all screening charges $S_{i}$ with $i\,<n-1,$
and operators $A_{1\text{ }}^{(m)}$ commute with all $S_{i}$, $i>1$. As a
result we derive that our matrix element can be represented as the ratio of
$sl(n)$ Selberg integrals
\begin{equation}
F_{(0,\dots  0,\mu)}^{(\lambda,0,\dots  ,0)}(a,P,P^{\prime})=\Omega_{\mu}(P)\Omega
_{\lambda}(P^{\prime})\frac{\left\langle \mathrm{J}_{\mu}^{(1/g)}[p_{k}+\eta]\mathrm{J}_{\lambda}^{(1/g)}[p_{-k}]\right\rangle _{Sel}^{sl(n)}%
}{\left\langle 1\right\rangle _{Sel}^{sl(n)}} \label{sin}%
\end{equation}
where $\eta=(a-Q)/b$, $g=-b^{2}$ and $\left\langle \dots \right\rangle _{Sel}^{sl(n)}$
denotes Selberg average, which can be defined in the following way \cite{Tarasov-Varchenko,Warnaar-An}. We define the functions
\begin{equation}
D(t^{(r)})=\prod\limits_{1\leq i<j\leq k_{r}}(t_{j}^{(r)}-t_{i}^{(r)}),\quad D(t^{(r)},t^{(r+1)})=\prod\limits_{i=1}^{k_{r}}\prod\limits_{j=1}^{k_{r+1}}
(t_{i}^{(r)}-t_{j}^{(r+1)}) \label{d}%
\end{equation}
and collection of contours $\mathcal{C(}n)\mathcal{=}C_{1}\times C_{2}%
\times\dots  C_{k_{1}+\dots +k_{n-1}}$, starting at the point $0$ and encircling the point $1$ and such that $C_{i}$ is between the interval $[0,1]$ and $C_{j}$ for $i<j$ as it is shown on fig. \ref{contours}.
Let $\mathcal{O}(t^{(1)}|\dots  |t^{(n-1)})$ be symmetric (with respect of $t_{a}^{(s)}\rightarrow$ $t_{b}^{(s)}$ ) single valued
function. We fix the branch of the function
\begin{equation}
G_{n}(t^{(1)}|\dots  |t^{(n-1)})=\prod\limits_{s=1}^{n-1}
D(t^{(r)})^{2g}\prod\limits_{i=1}^{k_{r}}
(t_{i}^{(r)})^{A_{r}}(1-t_{i}^{(r)})^{B_{r}}\prod\limits_{r=1}^{n-2}
D(t^{(r)},t^{(r+1)})^{-g} \label{gt}%
\end{equation}
where
\begin{align}
A_{r}  &  =-b(\mathcal{Q}+P,e_{r})=-1-b^{2}-x_{r}^{\prime}+x_{r+1}^{\prime
},\quad g=-b^{2}\nonumber\\
B_{1}  &  =B=-ba,\quad B_{2}=B_{3}=\dots  =B_{n-1}=0 \label{AB}%
\end{align}
in the following way: at the point, where all the variables are real and negative one
has $\arg(t_{i}^{(r)}-t_{j}^{(r+1)})=\pi$ for any $i,j$ and $\arg(t_{i}%
^{(r)}-t_{j}^{(r)})=0$ for $i<j$. Then
\begin{equation}
\left\langle \mathcal{O}\right\rangle _{Sel}^{sl(n)}=\int\limits_{\mathcal{C(}n)}
\mathcal{O}(t^{(1)}|\dots  |t^{(n-1)})G(t^{(1)}|\dots  |t^{(n-1)})d^{k_{1}}t^{(1)}%
\dots  d^{k_{n-1}}t^{(n-1)} \label{ev}%
\end{equation}
where $\ d^{k_{r}}t^{(r)}=dt_{1}^{(r)}\dots  dt_{k_{r}}^{(r)}$. As it will be
shown in Appendix \ref{Selberg-integral} the ratio (\ref{sin}) can be calculated exactly with the
expected result that $F_{(0,\dots  0,\mu)}^{(\lambda,0,\dots  ,0)}(a,P,P^{\prime})$ is
given by eq \eqref{men}.
\section{Factorization of the matrix elements}\label{Factor}
It follows from the previous consideration that the matrix element
\begin{equation}
F_{\scriptscriptstyle{\varnothing},\dots  ,\scriptscriptstyle{\varnothing},\nu^{\prime}}^{\nu,\scriptscriptstyle{\varnothing},\dots  ,\scriptscriptstyle{\varnothing}
}(a,P,P^{\prime})=\frac{\left\langle \Psi_{\nu,\scriptscriptstyle{\varnothing},\dots  ,\scriptscriptstyle{\varnothing},P^{\ast
}}|V_{a}|\Psi_{\scriptscriptstyle{\varnothing},\dots  ,\scriptscriptstyle{\varnothing},\nu^{\prime},P^{\prime}}\right\rangle
}{\left\langle \Psi_{\vec{0},P^{\ast}}|V_{a}|\Psi_{\vec{0},P^{\prime}%
}\right\rangle } \label{el}%
\end{equation}
is given by eq \eqref{men}. Here we use the properties of singular vectors of $W_{n}-$ algebra to show that all the matrix elements $F_{\vec{\nu}^{\prime}}^{\vec{\nu}}(a,P,P^{\prime})$ are given by the same equation. All proofs of the statements that will be done
here follow literally to the proofs in the ref. \cite{Alba:2010qc} and we will not reproduce them here. Some new phenomena that appears in $W_{n}$ case is related with consistency (bootstrap) relations very similar to that in the factorized scattering theories. We will not describe here in details the representation theory of $W_{n}$--algebra and consider only some special degenerate representations which are important for our consideration.

Consider the zeros of the function
\[
\Omega_{\nu_{n}}(P^{\prime})=(-b)^{|\nu_{n}|}\prod\limits_{m\neq n}\prod\limits_{(i,j)\in\nu_{n}}
(x_{mn}^{\prime}+ib+jb^{-1}).
\]
These zeros correspond to the conditions
\begin{equation}
(h_{m}-h_{n},P^{\prime})=(e_{m}+e_{m+1}+\dots +e_{n-1},P^{\prime})=(\hat{e}_{m,n-1},P^{\prime})=-rb-r^{\prime}/b \label{sing}%
\end{equation}
where $r>0,r^{\prime}>0\in\nu_{n}$. When we tend vector $P^{\prime}$ to these points our vector $|P^{\prime}\rangle_{\nu_{n}}$ (\ref{j}) tends to zero in the Fock space generated by the operators $A_{n}^{(-m)}$. However the state $|\Psi_{\scriptscriptstyle{\varnothing,\dots, \varnothing,\nu'}P^{\prime}}\rangle$ does not vanish being expressed in terms of generators $\mathrm{\hat{W}}_{k}^{(-n)}$ and operators $a_{-m}$ applied to the vacuum state $|\Theta_{P^{\prime}}\rangle$. Instead we have proved that the matrix elements \eqref{el} are given by eq \eqref{men} and do not vanish at these points.

The condition (\ref{sing}) means that for these $P^{\prime}$ in Verma module $|\Theta_{P^{\prime}}\rangle$ exists a singular vector $\chi_{\hat{P}^{\prime}}$ at the level $rr^{\prime}$
\begin{equation}
|\chi_{P^{\prime}}\rangle=\mathrm{D}_{r,r^{\prime}}|\Theta_{P^{\prime}}\rangle=\left(
(\mathrm{\hat{W}}_{n}^{(-1)})^{rr^{\prime}}+\dots \right)  |\Theta_{P^{\prime}}\rangle\label{sv}
\end{equation}
(where omitted terms have degree in $\mathrm{\hat{W}}_{n}^{(-1)}$ at most $rr^{\prime}-1$) such that $\mathrm{\hat{W}}_{l}^{(k)}|\chi_{\hat{P}^{\prime}}\rangle=0$ for any $k>0$. 
We introduce the weights $\omega_{i}=h_{1}+\dots  +h_{i}$, $(\omega_{i},e_{j})=\delta_{ij}$ of fundamental representations $\pi(\omega_{i})$ of $sl(n)$. Consider, for example, the condition $(\hat{e}_{n-1,n-1},P^{\prime})=(e_{n-1},P^{\prime})=-rb-r^{\prime}/b$. The solution to
this equation can be written as
\begin{equation}
P_{r,r^{\prime}}^{\prime}=-\omega_{n-1}(r_{n-1}b+r_{n-1}^{\prime}/b)+P_{\bot
}^{\prime} \label{n-1}
\end{equation}
where $(e_{n-1},P_{\bot}^{\prime})=0$. For example $P_{\bot}^{\prime}=P^{\prime}+h_{n}(P,e_{n-1})$. The corresponding singular vector $\chi$ can be parametrized by the vector  $P_{r,\hat{t}_{n-1}r^{\prime}}^{\prime}=P_{r,r^{\prime}}^{\prime}+e_{n-1}r^{\prime}b^{-1}$. Here the symbol $\hat{t}_{n-1}r^{\prime}$ denotes that we make the Weyl reflection of the vector $-\omega_{n-1}r^{\prime}/b$ with respect to the plane orthogonal to vector
$e_{n-1}$\footnote{We denote by symbol $\hat{t}_{k}$ the Weyl reflection with respect to the simple root $e_{k}$ and by symbol $\hat{s}_{k}$ the Weyl reflection with respect to the root $\hat{e}_{k,n-1}$.}. We note that solution of arbitrary equation $(\hat{e}_{m,n-1},P^{\prime})=-r_{m}b-r_{m}^{\prime}/b$ can be written in the the form (\ref{n-1})  with numbers $r_{m},$ $r_{m}^{\prime}$ and with vector $P_{\bot}^{\prime}=P^{\prime}+h_{n}(P^{\prime},\hat{e}_{m,n-1})$ which satisfies the condition $(\hat{e}_{m,n-1},P_{\bot(m)}^{\prime})=0$. The corresponding singular vector at the level $r_{m}r_{m}^{\prime}$ will be parametrized by the vector $P_{r_{m},\hat{s}_{m}r_{m}^{\prime}}^{\prime}=P_{r_{m},r_{m}^{\prime}}^{\prime}+\hat{e}_{m,n-1}r_{m}^{\prime}b^{-1}$.  It easy to check that $x_{m}^{\prime}(r_{m},r_{m}^{\prime})=x_{n}^{\prime}(r_{m},r_{m}^{\prime})-\left(  r_{m}b+r_{m}^{\prime}/b\right)$ 
and parameters $x_{i}^{\prime}(r_{m},\hat{s}_{m}r_{m}^{\prime})=(h_{i},P_{r_{m},r_{m}^{\prime}}^{\prime})$ are
\begin{align}
x_{n}^{\prime}(r_{m},\hat{s}_{m}r_{m}^{\prime})  &  =\frac{n-1}{n}(r_{m}b+r_{m}^{\prime}/b)-r_{m}^{\prime}/b+h_{n}P_{\bot}^{\prime}%
=x_{n}^{\prime}(r_{m},r_{m}^{\prime})-r_{m}^{\prime}/b\nonumber\\
x_{m}^{\prime}(r_{m},\hat{s}_{m}r_{m}^{\prime})  &  =-\frac{1}{n}\left(
r_{m}b+r_{m}^{\prime}/b\right)  +r_{m}^{\prime}/b+h_{m}P_{\bot}^{\prime}=x_{n}^{\prime}(r_{m},r_{m}^{\prime})-r_{m}b\nonumber\\
x_{i}^{\prime}(r_{m},\hat{s}_{m}r_{m}^{\prime})  &  =x_{i}^{\prime}(r_{m},r_{m}^{\prime}),\quad i\neq n,m. \label{xrr}%
\end{align}
For the further consideration we need to calculate the values
\begin{equation}
\mathrm{R}_{r_{m},r_{m}^{\prime}}(P,P_{r_{m},r_{m}^{\prime}}^{\prime},a)=
\frac{\left\langle \Theta_{P^{\ast}}|V_{a}|\mathrm{D}_{r_{m},r_{m}^{\prime}}\Theta_{P_{m}^{\prime}}\right\rangle }{\left\langle \Theta_{P^{\ast
}}|V_{a}|\Theta_{P_{m}^{\prime}}\right\rangle } \label{rat}%
\end{equation}
where $P_{m}^{\prime}=P_{r_{m},r_{m}^{\prime}}^{\prime}$ and normalization of
$\mathrm{D}_{r_{m},r_{m}^{\prime}}$ is defined by (\ref{sv}). We note that the numerator and the denominator (in usual normalization) in this ratio vanish when $P^{\prime}$ tends to $P_{r_{m},r_{m}^{\prime}}^{\prime},$ however their ratio is finite and well defined. This ratio can be calculated in two different ways. The first is based on the known matrix element (\ref{el}) and second on the calculations in conformal Toda theory. In the first way we can take in (\ref{el}) as $\nu^{\prime}$ the rectangular diagram of size $r_{m}r_{m}^{\prime}$ and $\nu=\varnothing$. 
For such diagram we have
$$
|\Psi_{\scriptscriptstyle{\varnothing},\dots,\scriptscriptstyle{\varnothing},\nu^{\prime},P_{m}^{\prime}}\rangle=
c_{^{r_{m}r_{m}^{\prime}}}\left(  \mathrm{D}_{r_{m},r_{m}^{\prime}}\right)|\Theta_{P_{m}^{\prime}}\rangle
$$ 
because as follows from the results of Feigin and Fuks \cite{Feigin-Fuks} the vector vanishing in the Fock space is the descendant of the
singular vector. In our case we have the singular vector at the level $r_{m}r_{m}^{\prime}$ and hence our vector
$|\Psi_{\scriptscriptstyle{\varnothing},\dots  ,\scriptscriptstyle{\varnothing},\nu^{\prime},P_{m}^{\prime}}\rangle$ is the singular vector at this level. To find
the constant $c_{^{r_{m}r_{m}^{\prime}}}$ we note that asymptotics in parameter $a$ of $\mathrm{R}_{r_{m},r_{m}^{\prime}}(P,P^{\prime},a)$ is
defined by the main term $(\mathrm{\hat{W}}_{n}^{(-1)})^{r_{m}r_{m}^{\prime}}$ in the operator $\mathrm{D}_{r_{m},r_{m}^{\prime}}$. It is easy to derive from the the Ward identities that this asymptotics is equal to $\left(  \frac{\hat{w}_{n}(a)}{n-1}\right)  ^{r_{m},r_{m}^{\prime}}$ where $\hat{w}_{n}(a)$ is the
eigenvalue of the operator $\mathrm{\hat{W}}_{n}^{(0)}$ i.e. the main singular term in the OPE of $\mathrm{\hat{W}}_{n}(z)$ with the field 
$V_{a}(z^{\prime})=\exp(a\omega_{1}\varphi(z^{\prime}))$
\begin{equation}
\mathrm{\hat{W}}_{n}(z)V_{a}(z^{\prime})=\frac{\hat{w}_{n}(a)}{(z-z^{\prime
})^{n}}V_{a}(z^{\prime})+O((z-z^{\prime})^{1-n}) \label{wa}%
\end{equation}
It follows immediately from eq \eqref{m} that the asymptotic of $\hat{w}_{n}(a)$
is equal ($(h_{i},h_{j})=\delta_{j}^{i}-\frac{1}{n}$) to
\begin{equation}
\hat{w}_{n}(a)\rightarrow(-)^{n}a^{n}\prod\limits_{i=1}^{n}
(\omega_{1}h_{i})=(-)^{n}a^{n}\frac{(n-1)}{n}(-)^{n-1}\left(  \frac{1}%
{n}\right)  ^{(n-1)}=-a^{n}(n-1)\left(  \frac{1}{n}\right)  ^{n} \label{as}%
\end{equation}
The exact expression for $\mathrm{R}_{r_{m},r_{m}^{\prime}}(P,P^{\prime},a)$
is given by eq \eqref{men}
\begin{multline}\label{Rr}
\mathrm{R}_{r_{m},r_{m}^{\prime}}(P,P_{r_{m},r_{m}^{\prime}}^{\prime},a)=\prod\limits_{j=1}^{n}\prod\limits_{s\in\nu^{\prime}}
\left(  Q-E_{\nu_{,}^{\prime}\scriptscriptstyle{\varnothing}_{j}}(x_{j}-x_{n}^{\prime}(r_{m}%
,r_{m}^{\prime})|s)-a/n)\right)  =\\=
\prod\limits_{j=1}^{n}\prod\limits_{k=0}^{r_{m}^{\prime}-1}\prod\limits_{l=0}^{r_{m}-1}
(x_{n}^{\prime}(r_{m},r_{m}^{\prime})-a/n-x_{i}-bl-k/b)
\end{multline}
and has the asymptotic $(-\frac{a}{n})^{nr_{m}r_{m}^{\prime}}$. 
Comparing it with the asymptotic of 
$$
\left(  \frac{\hat{w}_{n}(a)}{n-1}\right)  ^{r_{m}%
,r_{m}^{\prime}}=(-)^{r_{m},r_{m}^{\prime}}(\frac{a}{n})^{nr_{m}r_{m}^{\prime}},
$$ 
we find that $c_{^{_{r_{m}r_{m}^{\prime}}}}=(-)^{(n-1)r_{m}r_{m}^{\prime}}$. To have no sign factors we accept the factor 
$c_{^{_{r_{m}r_{m}^{\prime}}}}$ in the normalization of the operator $\mathrm{D}_{r,r^{\prime}}$ and the
vector 
$$
|\Psi_{P}^{(\bar{\nu})}\rangle=((-)^{(n-1)|\bar{\nu}|}((\mathrm{\hat{W}}%
_{n}^{(-1)})^{rr^{\prime}}+\dots ))|\Theta_{P}\rangle.
$$

To calculate the same value using conformal Toda theory we note that the state $|\Theta_{P}\rangle=|\Theta_{P}^{(w)}\rangle|\theta_{F}\rangle$ \ where
$|\theta_{F}\rangle$ is the Fock vacuum annihilated by operators $a_{k}$ with $k>0$ and $|\Theta_{P}^{(w)}\rangle$ is the state annihilated by all
$\mathrm{\hat{W}}_{l}^{(k)}$ ($2\leq l\leq n$) with $k>0$. The main objects in conformal Toda theory are the exponential fields $V_{\alpha}(z,\bar
{z})=e^{\alpha\varphi(z,\bar{z})}$ where $\varphi(z,\bar{z})$ is a Toda field. The primary states of $W_{n}\otimes\overline{W}_{n}-$ algebra $|\hat{\Theta
}_{P}\rangle$ are created (up to some factor) by the application of these fields to the Toda vacuum state $|\hat{\Theta}_{\scriptscriptstyle{\varnothing}}\rangle$ which satisfies the conditions $\mathrm{\hat{W}}_{l}^{(k)}|\hat{\Theta}_{\scriptscriptstyle{\varnothing}
}\rangle=\overline{\mathrm{\hat{W}}}_{l}^{(k)}|\hat{\Theta}_{\scriptscriptstyle{\varnothing}}%
\rangle=0$ for $k\geq-l+1$. It means that the state $|\hat{\Theta}_{P}%
\rangle=\lim_{z\rightarrow0}c(\alpha)V_{\alpha}(z,\overline{z})|\hat{\Theta
}_{\scriptscriptstyle{\varnothing}}\rangle$ ($P=\alpha-\mathcal{Q}$)$\ $\ and the state $\langle
\hat{\Theta}_{P^{\ast}}|=\lim_{z\rightarrow\infty}c(2\mathcal{Q}%
-\alpha)|z|^{2\Delta(\alpha)}\langle\hat{\Theta}_{\scriptscriptstyle{\varnothing}}|V_{2\mathcal{Q}%
-\alpha}(z,\bar{z})$. As the generators of $W_{n}-$ algebra commute with
generators of $\overline{W}_{n}$ we have in particular that for any $P$ and
$P^{\prime}$
\begin{equation}
\left(  \frac{\left\langle \Theta_{P^{\ast}}|V_{a}|\mathrm{D}_{r_{m}%
,r_{m}^{\prime}}\Theta_{P^{\prime}}\right\rangle }{\left\langle \Theta
_{P^{\ast}}|V_{a}|\Theta_{P^{\prime}}\right\rangle }\right)  ^{2}%
=\frac{\left\langle \hat{\Theta}_{P^{\ast}}|V_{a}|\mathrm{D}_{r_{m}%
,r_{m}^{\prime}}\mathrm{\bar{D}}_{\mathrm{\ }r_{m},r_{m}^{\prime}}\hat{\Theta
}_{P^{\prime}}\right\rangle }{\left\langle \Theta_{P^{\ast}}|V_{a}%
|\Theta_{P^{\prime}}\right\rangle }. \label{kv}%
\end{equation}
The denominator of this ratio is known explicitly \cite{Fateev:2008bm,Fateev:2007ab}. It is equal
$c(\alpha)c(2\mathcal{Q}-\alpha)C(2\mathcal{Q}-\alpha,a\omega_{1}%
,\alpha^{\prime})$ where $C(2\mathcal{Q}-\alpha,a\omega_{1},\alpha^{\prime})$
is the three point function in conformal Toda theory
\begin{equation}
C(2\mathcal{Q}-\alpha,a\omega_{1},\alpha^{\prime})=\frac{\mathrm{m}%
^{(P,\rho)/b}\mathrm{m}^{-(P^{\prime},\rho)/b}\mathrm{m}^{-a}(\Upsilon
(b))^{n-1}\Upsilon(a)\prod\limits_{e>0}
\Upsilon(-Pe)\Upsilon(P^{\prime}e)}{\prod\limits_{i,j}
\Upsilon(a/n+Ph_{i}-P^{\prime}h_{j})} \label{3p}%
\end{equation}
here $\mathrm{m}$ is non-essential function of $b$ and Toda cosmological constant. It is convenient to take $c(\alpha)=(\mathrm{m}^{-(P,\rho)/b}\prod\limits_{e>0}\Upsilon(Pe))^{-1}$. Then the field $c(\alpha)V_{\alpha}$ will be Weyl invariant and at 
$P^{\prime}=P_{r_{m},r_{m}^{\prime}}^{\prime}+\varepsilon$ where $|\varepsilon|\rightarrow0$
\begin{equation}
\mathrm{D}_{r_{m},r_{m}^{\prime}}\overline{\mathrm{D}}_{r_{m},r_{m}^{\prime}}|\hat{\Theta}_{P^{\prime}}\rangle=
(-)^{nr_{m}r_{m}^{\prime}}|\hat{\Theta}_{P^{^{\prime}}-rm^{\prime}\hat{e}_{mn}}\rangle+O(|\varepsilon|) \label{nv}%
\end{equation}
If we take into account eq \eqref{xrr} we derive that
\begin{equation}
\begin{aligned}
&  \left(  \mathrm{R}_{r_{m},r_{m}^{\prime}}(P,P_{r_{m},r_{m}^{\prime}}%
^{\prime},a)\right)  ^{2}\nonumber\\
&  =(-)^{nr_{m}r_{m}^{\prime}}\frac{\prod\limits_{i}
\Upsilon(a/n+x_{i}-x_{n}^{\prime}(r_{m},r_{m}^{\prime})+r_{m}b+r_{m}^{\prime
}/b)\Upsilon(a/n+x_{i}-x_{n}^{\prime}(r_{m},r_{m}^{\prime}))}{\prod\limits_{i}
\Upsilon(a/n+x_{i}-x_{n}^{\prime}(r_{m},r_{m}^{\prime})+r_{m}b)\Upsilon
(a/n+x_{i}-x_{n}^{\prime}(r_{m},r_{m}^{\prime})+r_{m}^{\prime}/b)} \label{rrr}%
\end{aligned}
\end{equation}
Function $\Upsilon(x)$ \cite{Zamolodchikov:1995aa} satisfies the functional relations $\Upsilon(x+b)=\gamma(bx)b^{1-2bx}\Upsilon(x),\Upsilon(x+b^{-1})=\gamma(x/b)b^{-1+2bx}%
\Upsilon(x),$ where $\gamma(x)=\frac{\Gamma(x)}{\Gamma(1-x)}$. So
\begin{multline}
\left(  \mathrm{R}_{r_{m},r_{m}^{\prime}}(P,P_{r_{m},r_{m}^{\prime}}^{\prime},a)\right)  ^{2}=\\=
(-)^{nr_{m}r_{m}^{\prime}}b^{2nr_{m}r_{m}^{\prime}}\prod\limits_{i}\prod\limits_{k=0}^{r_{m}^{\prime}-1}
\frac{\gamma(b^{-1}(a/n+x_{i}-x_{n}^{\prime}(r_{m},r_{m}^{\prime})+r_{m}b+k/b))}{\gamma\left(  b^{-1}(a/n+x_{i}-x_{n}^{\prime}(r_{m},r_{m}^{\prime})+r_{m}^{\prime}/b)\right)  }=\\=
\prod\limits_{i}\prod\limits_{k=0}^{r_{m}^{\prime}-1}
\prod\limits_{l=0}^{r_{m}-1}
(a/n+x_{i}-x_{n}^{\prime}(r_{m},r_{m}^{\prime})+bl+k/b)^{2} \label{RR}%
\end{multline}
It is easy to see that this expression is the square of the r.h.s. of (\ref{Rr}).
As we know the asymptotics of $\mathrm{R}_{r_{m},r_{m}^{\prime}}$ at $a\rightarrow\infty$ we can take the proper root in eq \eqref{RR}. We denote as
$\mathrm{R}_{r,r^{\prime}}^{(k)}(P,P^{\prime},a)$ the function
\[
\mathrm{R}_{r,r^{\prime}}^{(k)}(P,P^{\prime},a)=\prod\limits_{i}^{n}\prod\limits_{j=0}^{r^{\prime}-1}\prod\limits_{l=0}^{r-1}
(h_{k}P^{\prime}-a/n-h_{i}P-bl-j/b).
\]

With the function $\mathrm{R}_{r_{m},r_{m}^{\prime}}$ we can formulate the
factorization property for the functions $F_{\vec{\nu}}^{\vec{\lambda}}%
(P,P^{\prime},a)$ which states
\begin{equation}
F_{\vec{\nu}}^{\vec{\lambda}}(P,P_{r_{m},r_{m}^{\prime}}^{\prime}%
a)=\mathrm{R}_{r_{m},r_{m}^{\prime}}^{(n)}(P,P_{r_{m},r_{m}^{\prime}}^{\prime
},a)F_{\vec{\sigma}}^{\vec{\lambda}}(P,P_{r_{m},\hat{s}r_{m}^{\prime}}%
^{\prime},a) \label{fp}%
\end{equation}
here ``vectors'' $\vec{\lambda}=(\lambda^{(1)},\varnothing,\dots,\varnothing)$, $\nu=(\varnothing,\dots,\varnothing,\nu^{(n)})$ and  ``vector''
$\vec{\sigma}$ has only two non-zero components $\sigma^{(m)}$ and $\sigma^{(n)}$ where
$\sigma^{(n)}=(\nu_{1}^{(n)}-r_{m}^{\prime},\dots ,\nu_{r_{m}}^{(n)}%
-r_{m}^{\prime}),$ and $\sigma^{(m)}=(\nu_{r_{m}+1,}^{(n)}\nu_{r_{m}+2,}%
^{(n)}\dots )$. It is easy to check that this relation holds for the functions
(\ref{men}). Really this relation determines the functions $F_{\vec{\sigma}%
}^{\vec{\lambda}}(P,P^{\prime},a)$ with arbitrary partitions $\sigma^{(m)}$
and $\sigma^{(n)}$ (see ref. \cite{Alba:2010qc} for proof). An example of how the pair of the partitions $\sigma^{(m)}$ and $\sigma^{(n)}$ is defined for given $(r_{m},r_{m}')\in\nu^{(n)}$ is shown by the following picture
\begin{equation*}
  \begin{picture}(50,120)(90,-90)
    \Thicklines
    \unitlength 2.9pt 
   \rotatebox{270}{
    \put(0,20){\line(0,1){50}}
    \put(15,20){\line(0,1){35}}
    \put(0,20){\line(1,0){34}}
    \put(0,45){\line(1,0){15}}
    \put(0,70){\line(1,0){3}}
    \put(3,70){\line(0,-1){3}}
    \put(3,67){\line(1,0){3}}
    \put(6,67){\line(0,-1){4}}
    \put(6,63){\line(1,0){4}}
    \put(10,63){\line(0,-1){5}}
    \put(10,58){\line(1,0){3}}
    \put(13,58){\line(0,-1){3}}
    \put(13,55){\line(1,0){2}}
    \put(15,53){\line(1,0){2}}
    \put(17,53){\line(0,-1){3}}
    \put(17,50){\line(1,0){2}}
    \put(19,50){\line(0,-1){4}}
    \put(19,46){\line(1,0){3}}
    \put(22,46){\line(0,-1){6}}
    \put(22,40){\line(1,0){2}}
    \put(24,40){\line(0,-1){3}}
    \put(24,37){\line(1,0){2}}
    \put(26,37){\line(0,-1){6}}
    \put(26,31){\line(1,0){2}}
    \put(28,31){\line(0,-1){3}}
    \put(28,28){\line(1,0){2}}
    \put(30,28){\line(0,-1){4}}
    \put(30,24){\line(1,0){2}}
    \put(32,24){\line(0,-1){2}}
    \put(32,22){\line(1,0){2}}
    \put(34,22){\line(0,-1){2}}
    \put(11,16){\vector(1,0){4}}
    \put(4,16){\vector(-1,0){4}}
    \put(-3.8,36){\vector(0,1){8}}
    \put(-3.8,29){\vector(0,-1){8}}
    }
    \put(-18,-8){\mbox{$\sigma^{(n)}$}}
    \put(-40,-21){\mbox{$\sigma^{(m)}$}}
    \put(-56,-8.4){\mbox{$r_{m}$}}
    \put(-39.6,3){\mbox{$r_{m}'$}}
    \put(-69,-17){\mbox{$\nu^{(n)}\rightarrow$}}
    \end{picture}
\end{equation*}
Equation \eqref{fp} looks very similar to the equations for the formfactors of bound state (diagram $\nu^{(n)}$) in terms of formfactors of the particles (diagrams  $\sigma^{(m)}$ and $\sigma^{(n)}$) which form this bound state.
To underline this similarly we note that the conservation laws are fulfilled for this
fusion process. For example, it is easy to check for conserved charges $q^{(3)}(\nu,x)$ defined by \eqref{eig} that
\begin{multline}
  q^{(3)}(\nu^{(n)},x-x(r_{m},r'_{m}))=q^{(3)}(\sigma^{(n)},x-x(r_{m},-r'_{m}))+q^{(3)}(\sigma^{(m)},x+x(r_{m},-r'_{m}))+\\
  +q^{(3)}(\square_{r_{m},r'_{m}} ,x-x(r_{m},r'_{m})),
\end{multline}
where $\square_{r_{m},r'_{m}}$ is the rectangular diagram of size $r_{m}\times r'_{m}$ and $x(r,r')=-rb/2-r'b^{-1}/2$.

To proceed further we should impose two conditions (\ref{sing}) to the vector $P^{\prime}$. We note that if impose even $n-1$ conditions (\ref{sing}) to the vector $P^{\prime}$ the representation of $W_{n}-$ algebra will not be strongly degenerate. The strongly degenerate representations are specified by the vector $P^{\prime}$ such that  for all simple roots $(e_{i},P)=-r_{i}b-r_{i}^{\prime}b^{-1}$, where all numbers $r_{i},r_{i}^{\prime}$ are
positive. These representations contain $n!-1$ singular vectors. The representations specified by the vector $P^{\prime}$ satisfying $n-1$
condition (\ref{sing}) will have in general $2^{(n-1)}-1$ singular vectors.

\begin{figure}
\psfrag{a}{$|P_{r_{1},r_{1}';r_{2},r_{2}'}\rangle$}
\psfrag{b}{$r_{1}r_{1}'$}
\psfrag{c}{$r_{2}r_{2}'$}
\psfrag{d}{$(r_{1}-r_{2})r_{1}'+r_{2}r_{2}'$}
	\centering
	\includegraphics[width=.24\textwidth]{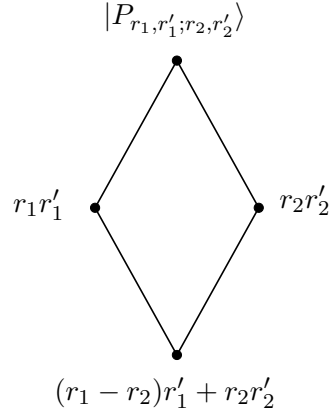}
        \caption{Embedding diagram for representation $P_{r_{1},r_{1}^{\prime};r_{2},r_{2}^{\prime}}^{\prime}$.}
	\label{embdiag}
\end{figure}

To make our consideration more clear and do not introduce the long notation we consider the case $n=3$. In this case we have two simple roots $e_{1}$ $e_{2}$ and the root $\hat{e}_{12}=e_{1}+e_{2}$. We also have $e_{1}=2\omega_{1}-\omega_{2},e_{2}=2\omega_{2}-\omega_{1}$. 
The solution of two equations \eqref{sing} is
\begin{equation}
P_{r_{1},r_{1}^{\prime};r_{2},r_{2}^{\prime}}^{\prime}=-\omega_{1}%
((r_{1}-r_{2})b+(r_{1}^{\prime}-r_{2}^{\prime})b^{-1})-\omega_{2}(r_{2}%
b+r_{2}^{\prime}b^{-1}) \label{or}%
\end{equation}
where $r_{2},r_{2}^{\prime}>0$ and $r_{1}-r_{2}>0$, $r_{1}^{\prime}-r_{2}^{\prime}<0$. In this case we will have in the Verma module two singular
vectors. The first at the level $r_{2}r_{2}^{\prime}$ parametrized by
\begin{equation}
P_{r_{1},r_{1}^{\prime};r_{2},\hat{s}_{2}r_{2}^{\prime}}^{\prime}%
=P_{r_{1},r_{1}^{\prime};r_{2},r_{2}^{\prime}}^{\prime}+e_{2}r_{2}^{\prime
}/b=-\omega_{1}((r_{1}-r_{2})b+r_{1}^{\prime}b^{-1})-\omega_{2}(r_{2}%
b-r_{2}^{\prime}b^{-1}) \label{fir}%
\end{equation}
and the second at the level $r_{1}r_{1}^{\prime}$ parametrized by
\begin{equation}
P_{r_{1},\hat{s}_{1}r_{1}^{\prime};r_{2},r_{2}^{\prime}}^{\prime}%
=P_{r_{1},r_{1}^{\prime};r_{2},r_{2}^{\prime}}^{\prime}+\hat{e}_{12}%
r_{1}^{\prime}/b=-\omega_{1}((r_{1}-r_{2})b-r_{1}^{\prime}b^{-1})-\omega
_{2}(r_{2}b+(r_{2}^{\prime}-r_{1}^{\prime})b^{-1}) \label{sec}%
\end{equation}
But the primary state parametrized by $P_{r_{1},r_{1}^{\prime};r_{2},\hat
{s}_{2}r_{2}^{\prime}}^{\prime}$ has a singular vector at the level
$(r_{1}-r_{2})r_{1}^{\prime}$ with $P^{\prime}$ equal
\[
P_{r_{1},\hat{s}_{2}\hat{s}_{1}r_{1}^{\prime};r_{2},\hat{s}_{2}r_{2}^{\prime}%
}^{\prime}=P_{r_{1},\hat{s}_{1}r_{1}^{\prime};r_{2},r_{2}^{\prime}}^{\prime
}+e_{1}r_{1}^{\prime}b^{-1}=-\omega_{1}((r_{1}-r_{2})b-r_{1}^{\prime}%
b^{-1})-\omega_{2}(r_{2}b+(r_{1}^{\prime}-r_{2}^{\prime})b^{-1})
\]
and primary state parametrized by $P_{r_{1},r_{1}^{\prime};r_{2},\hat{s}%
_{2}r_{2}^{\prime}}^{\prime}$ has the a singular vector at the level
$r_{2}(r_{2}^{\prime}-r_{1}^{\prime})$ with the same $P^{\prime}=P_{r_{1}%
,\hat{s}_{2}\hat{s}_{1}r_{1}^{\prime};r_{2},\hat{s}_{2}r_{2}^{\prime}}%
^{\prime}$. We note that $\hat{s}_{2}\hat{s}_{1}=\hat{t}_{1}$ and $\hat{s}%
_{2}=\hat{t}_{2},$ where $\hat{t}_{i}$ are the Weyl reflections with respect
to the planes orthogonal to roots $e_{i}$. So we have in the Verma module with
$P_{r_{1},r_{1}^{\prime};r_{2},r_{2}^{\prime}}^{\prime}$ three singular
vectors at the levels $r_{2}r_{2}^{\prime},r_{1}r_{1}^{\prime}$ and
$(r_{1}-r_{2})r_{1}^{\prime}+r_{2}r_{2}^{\prime}$ with embedding diagram shown
in fig \ref{embdiag}.

We see that there are two ways to arrive to the last point corresponding to
$P^{\prime}=P_{r_{1},\hat{t}_{1}r_{1}^{\prime};r_{2},\hat{t}_{2}r_{2}^{\prime
}}^{\prime}$. In the first way we derive from eq \eqref{fp}
\[
F_{\scriptscriptstyle{\varnothing},\scriptscriptstyle{\varnothing},\nu^{(3)}}^{\lambda^{(1)},\scriptscriptstyle{\varnothing},\scriptscriptstyle{\varnothing}
}(P,P_{r_{1},r_{1}^{\prime};r_{2},r_{2}^{\prime}}^{\prime}a)=\mathrm{R}%
_{r_{2},r_{2}^{\prime}}^{(3)}(P,P_{r_{1},r_{1}^{\prime};r_{2},r_{2}^{\prime}%
}^{\prime},a)F_{\scriptscriptstyle{\varnothing},\sigma^{(2)},\sigma^{(3)}}^{\lambda^{(1)}%
,\scriptscriptstyle{\varnothing},\scriptscriptstyle{\varnothing}}(P,P_{r_{1},r_{1}^{\prime};r_{2},\hat{t}_{2}r_{2}%
^{\prime}}^{\prime},a)
\]
where $\sigma^{(3)}=(\nu_{1}^{(3)}-r_{2}^{\prime},\dots ,\nu_{r_{2}}^{(3)}-r_{2}^{\prime})$, $\sigma^{(2)}=(\nu_{r_{2}+1,}^{(3)}\nu_{r_{2}+2,}^{(3)}\dots )$ after the first step and after the second step
\begin{align}
F_{\scriptscriptstyle{\varnothing},\scriptscriptstyle{\varnothing},\nu^{(3)}}^{\lambda^{(1)},\scriptscriptstyle{\varnothing},\scriptscriptstyle{\varnothing}
}(P,P_{r_{1},r_{1}^{\prime};r_{2},r_{2}^{\prime}}^{\prime}a)  &
=\mathrm{R}_{r_{2},r_{2}^{\prime}}^{(3)}(P,P_{r_{1},r_{1}^{\prime};r_{2}%
,r_{2}^{\prime}}^{\prime},a)\mathrm{R}_{r_{1}-r_{2},r_{1}^{\prime}}%
^{(2)}(P,P_{r_{1},r_{1}^{\prime};r_{2},\hat{t}_{2}r_{2}^{\prime}}^{\prime
},a)\label{1w}\\
&  \times F_{\kappa^{(1)},\kappa^{(2)},\kappa^{(3)}}^{\lambda^{(1)}%
,\scriptscriptstyle{\varnothing},\scriptscriptstyle{\varnothing}}(P,P_{r_{1},\hat{t}_{1}r_{1}^{\prime};r_{2},\hat{t}%
_{2}r_{2}^{\prime}}^{\prime},a)
\end{align}
where $\kappa^{(3)}=\sigma^{(3)}$, $\kappa^{(2)}=
(\nu_{r_{2}+1}^{(3)}-r_{2}^{\prime}+r_{1}^{\prime},\dots ,\nu_{r_{1}}^{(3)}-r_{2}^{\prime}+r_{1}^{\prime})$ and
$\kappa^{(1)}=(\nu_{r_{1}+1}^{(3)},\nu_{r_{1}+2}^{(3)},\dots )$. It is convenient to represent it by the following picture:
\begin{equation*}
  \begin{picture}(0,150)(90,-110)
    \Thicklines
    \unitlength 2.4pt 
   \rotatebox{270}{
    \put(0,5){\line(0,1){65}}
    \put(15,5){\line(0,1){50}}
    \put(30,5){\line(0,1){20}}
    \put(15,20){\line(1,0){15}}
    \put(0,5){\line(1,0){45}}
    \put(0,45){\line(1,0){15}}
    \put(0,70){\line(1,0){3}}
    \put(3,70){\line(0,-1){3}}
    \put(3,67){\line(1,0){3}}
    \put(6,67){\line(0,-1){4}}
    \put(6,63){\line(1,0){4}}
    \put(10,63){\line(0,-1){5}}
    \put(10,58){\line(1,0){3}}
    \put(13,58){\line(0,-1){3}}
    \put(13,55){\line(1,0){2}}
    \put(15,53){\line(1,0){2}}
    \put(17,53){\line(0,-1){3}}
    \put(17,50){\line(1,0){2}}
    \put(19,50){\line(0,-1){4}}
    \put(19,46){\line(1,0){3}}
    \put(22,46){\line(0,-1){6}}
    \put(22,40){\line(1,0){2}}
    \put(24,40){\line(0,-1){3}}
    \put(24,37){\line(1,0){2}}
    \put(26,37){\line(0,-1){6}}
    \put(26,31){\line(1,0){2}}
    \put(28,31){\line(0,-1){3}}
    \put(28,28){\line(1,0){2}}
    \put(30,28){\line(0,-1){4}}
    \put(30,24){\line(1,0){2}}
    \put(32,24){\line(0,-1){2}}
    \put(32,22){\line(1,0){2}}
    \put(34,22){\line(0,-1){2}}
    \put(34,20){\line(1,0){3}}
    \put(37,20){\line(0,-1){2}}
    \put(37,18){\line(1,0){3}}
    \put(40,18){\line(0,-1){3}}
    \put(40,15){\line(1,0){3}}
    \put(43,15){\line(0,-1){3}}
    \put(43,12){\line(1,0){2}}
    \put(45,12){\line(0,-1){7}}
    \put(11,2){\vector(1,0){4}}
    \put(4,2){\vector(-1,0){4}}
    \put(19,-5){\vector(1,0){10}}
    \put(10,-5){\vector(-1,0){10}}
    \put(-8.8,30){\vector(0,1){14}}
    \put(-8.8,19){\vector(0,-1){14}}
    \put(-3.8,15){\vector(0,1){4}}
    \put(-3.8,9){\vector(0,-1){4}}}
    \put(-18,-8){\mbox{$\kappa^{(1)}$}}
    \put(-41,-23){\mbox{$\kappa^{(2)}$}}
    \put(-60,-39){\mbox{$\kappa^{(3)}$}}
    \put(-70,-8.4){\mbox{$r_{2}$}}
    \put(-76,-15.4){\mbox{$r_{1}$}}
    \put(-47.6,8){\mbox{$r_{2}'$}}
    \put(-59.6,3){\mbox{$r_{1}'$}}
    \put(-95,-23){\mbox{$\nu^{(3)}\rightarrow$}}
    \end{picture}
\end{equation*}

Taking the second way at the first step we have
\[
F_{\scriptscriptstyle{\varnothing},\scriptscriptstyle{\varnothing},\nu^{(3)}}^{\lambda^{(1)},\scriptscriptstyle{\varnothing},\scriptscriptstyle{\varnothing}
}(P,P_{r_{1},r_{1}^{\prime};r_{2},r_{2}^{\prime}}^{\prime}a)=\mathrm{R}%
_{r_{1},r_{1}^{\prime}}^{(3)}(P,P_{r_{1},r_{1}^{\prime};r_{2},r_{2}^{\prime}%
}^{\prime},a)F_{\mu^{(1)}\scriptscriptstyle{\varnothing},\mu^{(3)}}^{\lambda^{(1)},\scriptscriptstyle{\varnothing}
,\scriptscriptstyle{\varnothing}}(P,P_{r_{1},\hat{s}_{1}r_{1}^{\prime};r_{2},r_{2}^{\prime}}%
^{\prime},a)
\]
where $\mu^{(3)}=(\nu_{1}^{(3)}-r_{1}^{\prime},\dots ,\nu_{r\quad}^{(3)}%
-r_{1}^{\prime})$, $\mu^{(1)}=\kappa^{(1)}$ and after the second step
\begin{multline}
F_{\scriptscriptstyle{\varnothing},\scriptscriptstyle{\varnothing},\nu^{(3)}}^{\lambda^{(1)},\scriptscriptstyle{\varnothing},\scriptscriptstyle{\varnothing}
}(P,P_{r_{1},r_{1}^{\prime};r_{2},r_{2}^{\prime}}^{\prime}a) 
=\mathrm{R}_{r_{1},r_{1}^{\prime}}^{(3)}(P,P_{r_{1},r_{1}^{\prime};r_{2}%
,r_{2}^{\prime}}^{\prime},a)\mathrm{R}_{r_{2},r_{2}^{\prime}-r_{1}^{\prime}%
}^{(3)}(P,P_{r_{1},\hat{s}_{1}r_{1}^{\prime};r_{2},r_{2}^{\prime}}^{\prime
},a)\label{2w}\times\\\times F_{\kappa^{(1)},\kappa^{(2)},\kappa^{(3)}}^{\lambda^{(1)}%
,\scriptscriptstyle{\varnothing},\scriptscriptstyle{\varnothing}}(P,P_{r_{1},\hat{t}_{1}r_{1}^{\prime};r_{2},\hat{t}%
_{2}r_{2}^{\prime}}^{\prime},a),
\end{multline}
which is equivalent to the picture
\begin{equation*}
  \begin{picture}(0,150)(90,-110)
    \Thicklines
    \unitlength 2.4pt 
   \rotatebox{270}{
    \put(0,5){\line(0,1){65}}
    \put(15,20){\line(0,1){35}}
    \put(30,5){\line(0,1){20}}
    \put(0,20){\line(1,0){30}}
    \put(0,5){\line(1,0){45}}
    \put(0,45){\line(1,0){15}}
    \put(0,70){\line(1,0){3}}
    \put(3,70){\line(0,-1){3}}
    \put(3,67){\line(1,0){3}}
    \put(6,67){\line(0,-1){4}}
    \put(6,63){\line(1,0){4}}
    \put(10,63){\line(0,-1){5}}
    \put(10,58){\line(1,0){3}}
    \put(13,58){\line(0,-1){3}}
    \put(13,55){\line(1,0){2}}
    \put(15,53){\line(1,0){2}}
    \put(17,53){\line(0,-1){3}}
    \put(17,50){\line(1,0){2}}
    \put(19,50){\line(0,-1){4}}
    \put(19,46){\line(1,0){3}}
    \put(22,46){\line(0,-1){6}}
    \put(22,40){\line(1,0){2}}
    \put(24,40){\line(0,-1){3}}
    \put(24,37){\line(1,0){2}}
    \put(26,37){\line(0,-1){6}}
    \put(26,31){\line(1,0){2}}
    \put(28,31){\line(0,-1){3}}
    \put(28,28){\line(1,0){2}}
    \put(30,28){\line(0,-1){4}}
    \put(30,24){\line(1,0){2}}
    \put(32,24){\line(0,-1){2}}
    \put(32,22){\line(1,0){2}}
    \put(34,22){\line(0,-1){2}}
    \put(34,20){\line(1,0){3}}
    \put(37,20){\line(0,-1){2}}
    \put(37,18){\line(1,0){3}}
    \put(40,18){\line(0,-1){3}}
    \put(40,15){\line(1,0){3}}
    \put(43,15){\line(0,-1){3}}
    \put(43,12){\line(1,0){2}}
    \put(45,12){\line(0,-1){7}}
    \put(11,2){\vector(1,0){4}}
    \put(4,2){\vector(-1,0){4}}
    \put(19,-5){\vector(1,0){10}}
    \put(10,-5){\vector(-1,0){10}}
    \put(-8.8,30){\vector(0,1){14}}
    \put(-8.8,19){\vector(0,-1){14}}
    \put(-3.8,15){\vector(0,1){4}}
    \put(-3.8,9){\vector(0,-1){4}}}
    \put(-18,-8){\mbox{$\kappa^{(1)}$}}
    \put(-41,-23){\mbox{$\kappa^{(2)}$}}
    \put(-60,-39){\mbox{$\kappa^{(3)}$}}
    \put(-70,-8.4){\mbox{$r_{2}$}}
    \put(-76,-15.4){\mbox{$r_{1}$}}
    \put(-47.6,8){\mbox{$r_{2}'$}}
    \put(-59.6,3){\mbox{$r_{1}'$}}
    \put(-95,-23){\mbox{$\nu^{(3)}\rightarrow$}}    
    \end{picture}
\end{equation*}
It follows from eqs \eqref{1w} and \eqref{2w} that they are consistent if%
\begin{multline}\label{cc}
\mathrm{R}_{r_{2},r_{2}^{\prime}}^{(3)}(P,P_{r_{1},r_{1}^{\prime}%
;r_{2},r_{2}^{\prime}}^{\prime},a)\mathrm{R}_{r_{1}-r_{2},r_{1}^{\prime}%
}^{(2)}(P,P_{r_{1},r_{1}^{\prime};r_{2},\hat{t}_{2}r_{2}^{\prime}}^{\prime
},a)=\\
=\mathrm{R}_{r_{1},r_{1}^{\prime}}^{(3)}(P,P_{r_{1},r_{1}^{\prime}%
;r_{2},r_{2}^{\prime}}^{\prime},a)\mathrm{R}_{r_{2},r_{2}^{\prime}%
-r_{1}^{\prime}}^{(3)}(P,P_{r_{1},\hat{s}_{1}r_{1}^{\prime};r_{2}%
,r_{2}^{\prime}}^{\prime},a)
\end{multline}
It is easy to check that this equation as well as eqs \eqref{1w}--\eqref{2w} with functions $F_{\vec{\mu}}^{\vec{\lambda}}$ defined by eq \eqref{melm} are satisfied. The factorization equations \eqref{1w} and \eqref{2w} can be considered as equation for functions 
$F_{\kappa^{(1)},\kappa^{(2)},\kappa^{(3)}}^{\lambda^{(1)},\scriptscriptstyle{\varnothing},\scriptscriptstyle{\varnothing}}(P,P^{\prime},a)$. 
It can be proved that these equations have unique solution defined by eq \eqref{melm}. Having these functions we can develop the same procedure with partition $\lambda^{(1)}$ to derive all amplitudes $F_{\vec{\mu}}^{\vec{\lambda}}$.

\begin{figure}
	\centering
	\includegraphics[width=.21\textwidth]{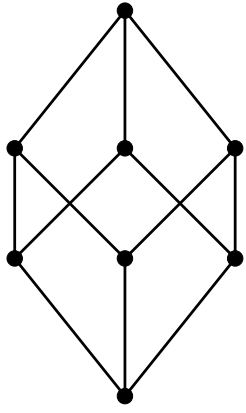}
        \caption{Embedding diagram for $sl(4)$.}
	\label{embdiag2}
\end{figure}

For arbitrary $n$ the number of singular vectors in the Verma module corresponding to the vector $P^{\prime}$ satisfying the $l$ equations (\ref{sing})
\begin{equation}
\hat{e}_{m_{i},n-1}P^{\prime}=-r_{m_{i}}b-r'_{m_{i}}b^{-1},\quad i=1\dots l, \label{Pc}
\end{equation}
depends on the position of points $m_{i}$ ($m_{i}$ form increasing sequence) on the Dynkin diagram of the Lie algebra $sl\left(n\right)$. We can delete
all nodes on this diagram except the nodes having numbers $m_{i}$. The resulting diagram will correspond to the Lie algebra 
$$
sl\left(  k_{1}\right)\otimes sl(k_{2})\otimes\dots \otimes sl(k_{j})
$$ 
where $\sum_{i=1}^{j}k_{i}=l+j$. The number of singular vectors will be in general case $N_{s}=\sum_{i=1}^{j}(2^{k_{j}-1}-1)$. 
The imbedding diagram of singular vectors for each of the Lie algebra $sl\left(  k\right)  $ has $k-1$ lines outgoing from starting point which enter to $k-1$ points corresponding to the first generation of the singular vectors. From each of these points starts $k-2$ lines which enter to $C_{k-1}^{2}$ points corresponding to the second generation of the singular vectors. From each of these points starts $k-3$ lines which end at $C_{k-1}^{3}$ points and so on. To the last point enter $k-1$ lines. The imbedding diagram for the Lie algebra $sl\left(  4\right)  $ is drown in fig \ref{embdiag2}. 
We see that at each point corresponding to the second generation we will have one consistency condition and in the last point we have three consistency condition. For arbitrary $k$ the number of these conditions will be $N_{c}=(n-1)\left(  n-2\right)  2^{n-4}$. It can be checked that all these conditions are satisfied.

With vectors $P^{\prime}$ satisfying eq \eqref{Pc} we can reproduce using factorization property the functions $F_{\vec{\sigma}}^{\lambda^{(1)}%
,\scriptscriptstyle{\varnothing},\dots  ,\scriptscriptstyle{\varnothing}}(P,P^{\prime},a)$ starting from known
function$\ F_{\scriptscriptstyle{\varnothing},\dots  ,\scriptscriptstyle{\varnothing},\nu^{(n)}}^{\lambda^{(1)},\scriptscriptstyle{\varnothing}
,\dots  ,\scriptscriptstyle{\varnothing}}(P,P^{\prime},a)$. The \textquotedblleft
vector\textquotedblright\ $\vec{\sigma}$ is specified by non-empty partitions $\sigma^{(m_{i})}$ where numbers $m_{i}$ correspond the position of point $m_{i}$ on the Dynkin diagram of $sl\left(  n\right)  $ and the partition $\sigma^{(n)}$. Then if we impose $n-1$ condition (\ref{Pc}) we can reproduce the functions $F_{\vec{\sigma}}^{\lambda^{(1)},\scriptscriptstyle{\varnothing},\dots  ,\scriptscriptstyle{\varnothing}}(P,P^{\prime},a)$ 
with $\vec{\sigma}=(\sigma^{(1)},\dots  ,\sigma^{(n)})$. Proceeding in the same way with the vector $P$ and partitions $\vec{\lambda}$ we
can reproduce from these functions all functions $F_{\vec{\sigma}}^{\vec{\lambda}}(P,P^{\prime},a)$ defined by eq (\ref{men}).

Thus we constructed the basis of states $|\Psi_{\vec{\nu},P}\rangle$ and primary operators \eqref{vertex} such that the matrix elements \eqref{me} are given by factorized expression \eqref{men}. The norms of these states $N(\vec{\nu},P)=\langle \Psi_{\vec{\nu},P}|\Psi_{\vec{\nu},P}\rangle$ can be obtained from \eqref{men} by setting $a=0$, $\vec{\nu}=\vec{\nu}'$ and $P=-P'$ i.e.
\begin{equation}
   N(\vec{\nu},P)=F_{\vec{\nu}}^{\vec{\nu}}(0,-P,P).
\end{equation}
Then the combinatorial expansion of the function \eqref{conformal-block-refined} takes the form
\begin{multline}\label{conformal-block-refined-explicit-coefficient}
   \mathbb{Z}_{\vec{j}}=
   \sum_{\vec{\nu}_{1},\dots,\vec{\nu}_{k-3}}
   N^{-1}(\vec{\nu}_{1},P_{1})\dots N^{-1}(\vec{\nu}_{k-3},P_{k-3})\times\\\times
   F^{\vec{\nu}_{1}}_{\scriptscriptstyle{\varnothing}}(a_{2},P_{1},P)
   F^{\vec{\nu}_{2}}_{\vec{\nu}_{1}}(a_{3},P_{2},P_{1})
   \dots
   F^{\vec{\nu}_{k-3}}_{\vec{\nu}_{k-4}}(a_{k-2},P_{k-3},P_{k-4})
   F^{\scriptscriptstyle{\varnothing}}_{\vec{\nu}_{k-3}}(a_{k-1},\hat{P},P_{k-3}),
\end{multline}
where $P=\alpha_{1}-\mathcal{Q}$ and $\hat{P}=\alpha_{k}-\mathcal{Q}$.
\section{Classical Integrability}
Here we consider the classical reduction of our integrable system. In the classical limit we define the fields
\begin{equation}
\mathsf{v}\rightarrow Q\phi^{\prime},\quad\mathsf{u}\rightarrow Q^{2}T,\quad\mathsf{u}_{k}\rightarrow Q^{k}\mathrm{\hat{W}}_{k},\quad\left[
,\right]  \rightarrow\frac{2\pi i}{Q^{2}}\left\{  ,\right\} , \label{pb}%
\end{equation}
and the operator $\mathrm{D}$: $\mathrm{D}F=i(F_{+}-F_{-})$, where $F_{+}$ and $F_{-}$ are the positive and negative frequency part of $F$. Then the densities of the first three integrals can be written as $\ I_{n,\left(
k\right)  }^{(cl)}=\frac{1}{2\pi}\int_{0}^{2\pi}G_{k}^{(n)}dx$
\begin{align}
G_{2}^{(n)}  &  =\mathsf{u}-\mathsf{v}^{2}/2,\nonumber\\
G_{3}^{(n)}  &  =\frac{n}{2i}\mathsf{v}\mathrm{D}\mathsf{v}+\frac{2}{\sqrt{n}%
}\mathsf{uv}-\frac{\mathsf{v}^{3}}{3\sqrt{n}}+i\mathsf{u}_{3}\nonumber\\
G_{4}^{(n)}  &  =-\frac{(1+n^{2})}{2}(\partial\mathsf{v)}^{2}+\frac
{(6-n)\mathsf{u}^{2}}{n}+\frac{3}{2i}\sqrt{n}(2\mathsf{u-v}^{2})\mathrm{D}%
\mathsf{v\mathsf{-}\frac{1}{4n}\mathsf{\mathsf{v}^{4}}+}\frac{3i}{\sqrt{n}%
}\mathsf{u}_{3}\mathsf{v+u}_{4}, \label{g23}%
\end{align}
where the terms $\mathsf{u}_{3}$ and $\mathsf{u}_{4}$ appear only for $n>2$ and $n>3$ respectively. The Poisson brackets for the fields $\mathsf{u}$, $\mathsf{u}_{3},\mathsf{u}_{4}\dots $ are the standard Gel'fand-Dikii brackets and can be found in refs \cite{Gelfand-1976,Drinfeld-Sokolov,Fateev:1991aw}. The Poisson bracket for the field $\mathsf{v}$ is: $\left\{  \mathsf{v(x)},\mathsf{v(y)}\right\}=-\delta^{\prime}(x-y)$.

The field $\mathsf{u}_{3}$ transforms as the primary field of spin $3$. It means that its Poisson bracket with $\mathsf{u}$ has canonical form
\[
\left\{  \mathsf{u(x),u}_{3}(y)\right\}  =(\delta^{\prime}(x-y)(\mathsf{u}%
_{3}(y)+2\mathsf{u}_{3}(x)).
\]
The field $\mathsf{u}_{4}$ is not ``primary''. In quantum case we can represent the field $\mathrm{\hat{W}}_{4}$ in the form
\begin{equation}
\mathrm{\hat{W}}_{4}=\widetilde{\mathrm{W}}_{4}+\frac{(n-2)(n-3)}{2}\left(
\frac{(\frac{5}{n}+3(5-3n)Q^{2})}{17+5n(n^{2}-1)Q^{2}}\Lambda+\frac{3nQ^{2}}{10}\partial^{2}T\right),  \label{w4}
\end{equation}
where $\Lambda=_{\times}^{\times}T$ $T_{\times}^{\times}-\frac{3}{10}\partial^{2}T$ (here symbol $_{\times}^{\times}\dots $\ $_{\times}^{\times}$
denotes regularized product) and $\widetilde{\mathrm{W}}_{4}$ is a primary field of spin four. In the semiclassical limit $Q\rightarrow\infty$ we find that
\[
\mathsf{u}_{4}=\widetilde{\mathsf{u}}_{4}+\frac{(n-2)(n-3)}{2}\left(
\frac{3(5-3n)}{5n(n^{2}-1)}\mathsf{u}^{2}+\frac{3n}{10}\partial^{2}%
\mathsf{u}\right).
\]
With the Gel'fand-Dikii brackets we can write the equation of motion for the functions $\mathsf{v,u},\mathsf{u}_{i}$. It is however easier to use the
bosonization procedure described above and express the integrals $I_{n,\left(  k\right)  }^{(cl)}$ in terms of the field $\phi_{i}=Q\Phi_{i}$
which have simple Poisson brackets $\left\{  \phi_{i}(x),\phi_{j}(y)\right\}=-\delta_{j}^{i}\delta^{\prime}(x-y)$. The densities $\mathrm{G}_{k}^{(n)}$
of the first three integrals in these variables have a form (we put for simplicity $x_{i}=0$)
\begin{align}
\mathrm{G}_{2}^{(n)}  &  =-\sum_{j=1}^{n}\phi_{j}^{2},\nonumber\\
\mathrm{G}_{3}^{(n)}  &  =\frac{1}{2i}\hat{\phi}\mathrm{D}\hat{\phi}+\sum_{j>k}^{n}\phi_{j}\partial\phi_{k}-\frac{1}{3}\sum_{j=1}^{n}\phi_{j}^{3},\nonumber\\
\mathrm{G}_{4}^{(n)}  &  =-\frac{n\partial\hat{\phi}\partial\hat{\phi}}{2}+\sum_{j=1}^{n}\frac{(n-1)}{2}\partial\phi_{j}\partial\phi_{j}-\frac
{\phi_{j}^{4}}{4}-\frac{1}{i}(\phi_{j}^{2}-(n+1-2j)\partial\phi_{j})\mathrm{D}\hat{\phi},\nonumber\\
&  +\sum_{k=1}^{n-1}(n-2k)\sum_{l=1}^{n-k-1}\partial\phi_{l}\partial\phi_{l+k}+\sum_{j=1}^{n}\sum_{k=1}^{n}\phi_{k}^{2}\sigma\lbrack j-k]\partial
\phi_{j}, \label{sgn}
\end{align}
where $\hat{\phi}=\sum_{i}^{n}\phi_{i}$ and $\sigma\lbrack i-j]=1$ if $i>j,$
$0$ if $i=j$ and $-1$ if $i\,<j$.

The equations of motion corresponding to the dynamics with Hamiltonian  $\mathcal{H}_{3}=i\ I_{n,\left(  3\right)  }^{(cl)}$ ($\partial_{t}\phi_{j}=\left\{  \phi_{j},\mathcal{H}_{3}\right\}  $) have a simple form
\begin{equation}\label{em}
\frac{1}{i}\partial_{t}\phi_{j}+\frac{1}{i}\mathrm{D}\hat{\phi}+\sum_{k=1}^{n}\sigma\lbrack j-k]\partial^{2}_{x}\phi_{k}-2\phi_{j}\partial_{x}\phi_{j}=0.
\end{equation}
They admit the ``reality'' condition $\phi_{j}^{\ast}=-\phi_{n+1-j}$. To study these equations we consider first the case $n=2$. In this case we have
\begin{equation}
\begin{aligned}
&  \frac{1}{i}\partial_{t}\phi_{1}+\frac{1}{i}\mathrm{D}(\phi_{1}+\phi
_{2})_{x}-\partial_{x}^{2}\phi_{2}-2\phi_{1}\partial_{x}\phi_{1}=0\\
&  \frac{1}{i}\partial_{t}\phi_{2}+\frac{1}{i}\mathrm{D}(\phi_{1}+\phi
_{2})_{x}+\partial_{x}^{2}\phi_{1}-2\phi_{2}\partial_{x}\phi_{2}=0. \label{sl2}%
\end{aligned}
\end{equation}
The reality condition can be written as $\phi_{1}=i\psi^{\ast},\phi_{2}=i\psi$. In terms of function $\psi$ these equations can be written as
\begin{equation}
\partial_{t}\psi+2\operatorname{Re}\mathrm{D}\psi_{x}+i\partial_{x}^{2}%
\psi^{\ast}+2\psi\partial_{x}\psi=0.\label{eq2}%
\end{equation}
This equation (up to the substitution $t\rightarrow2t$) coincides with the equation derived in \cite{Alba:2010qc}. For the functions $\psi$ analytical in the upper half plane this equation due to relation $\mathrm{D}(\psi+\psi^{\ast})=i(\psi_{x}-\psi^{*}_{x})$ reduces to the complex Burgers equation
\begin{equation}
\partial_{t}\psi+i\partial_{x}^{2}\psi+2\psi\partial_{x}\psi=0, \label{B}%
\end{equation}
which can be linearized by Cole-Hopf substitution $\psi=i(\log\theta)_{x}$%
\begin{equation}
\partial_{t}\theta+i\theta_{xx}=0. \label{ch}%
\end{equation}

We note that for all $n$\ eqs \eqref{em} can be reduced to the complex Burgers equation if we put $\phi_{2}=\phi_{3}=\dots =\phi_{n-1}=0$ and functions $\phi_{1}=i\psi^{*}$, $\phi_{n}=i\psi$ with $\psi$ analytical in the upper half plane.

If we take equation with $j=1$ and the difference between equations $j+1$ and $j$ in \eqref{em} we obtain
\begin{equation}
\begin{aligned}
&\frac{1}{i}\partial_{t}\phi_{1}+\frac{1}{i}\mathrm{D}\hat{\phi}_{x}-\sum
_{j=2}^{n}\partial^{2}\phi_{j}-2\phi_{1}\partial_{x}\phi_{1}=0,\\
&\frac{1}{i}\partial_{t}(\phi_{j+1}-\phi_{j})+\partial^{2}(\phi_{j}+\phi
_{j+1})-\partial_{x}(\phi_{j+1}^{2}-\phi_{j}^{2})=0
\end{aligned}
\end{equation}
In the limit $n\rightarrow\infty$ we can define variable $y=j/n,$ denote $\phi_{i}=\mathrm{\chi}_{j/n}$ and introduce the field 
$\mathrm{\rho}(x,y,t)=n\mathrm{\chi}_{i/n}(x,nt)$, then this system of equations can be
written as
\begin{equation}
\begin{aligned}
&\frac{1}{i}\partial_{y}\partial_{t}\mathrm{\rho}(x,y,t)+2\partial_{x}%
^{2}\mathrm{\rho}(x,y,t)-\partial_{x}\partial_{y}\mathrm{\rho}^{2}(x,y,t)=0,\\
&\frac{1}{i}\partial_{t}\mathrm{\rho}(x,0,t)+\int_{0}^{1}(\frac{1}%
{i}\mathrm{D\rho}(x,y,t)_{x}-\rho(x,y,t)_{xx})dy-\partial_{x}\mathrm{\rho}^{2}(x,0,t)=0.
\end{aligned}
\end{equation}
These equations have a simple stationary solution
\[
\mathrm{\rho}(x,y)=(1/2-y)\cot(x/2+i\,\textrm{sign}(y-1/2)\eta)
\]
where $\eta>0$. It is easy to check that for all $n$ the functions 
$$
\phi_{j}=(\frac{n+1}{2}-i)\cot(x/2+i\,\textrm{sign}(j-\frac{n+1}{2})\eta)
$$ 
give the stationary solution of eqs \eqref{em}.

We can consider dynamics generated by the next Hamiltonian $\mathcal{H}%
_{4}=\ I_{n,\left(  4\right)  }^{(cl)}$ ($\partial_{t_{2}}\phi_{i}=\left\{
\phi_{i},\mathcal{H}_{4}\right\}  $). The corresponding equation looks rather
complicated. However, if we take the same reduction: $\phi_{1}=i\psi^{\ast
}$ , $\phi_{n}=i\psi$ an all other $\phi_{j}=0$, the equation for $\psi$
analytical in upper half plane has a form
\begin{equation}\label{h4}
\partial_{t_{2}}\psi+\partial_{x}^{3}\psi-3i\partial_{x}(\psi\partial_{x}\psi)-\partial_{x}(\psi)^{3}=0 
\end{equation}
This equation also can be linearized by Cole-Hopf substitution (\ref{ch})
\[
\partial_{t_{2}}\theta+\theta_{xxx}=0
\]
It looks reasonable to think that for higher Hamiltonians $\mathcal{H}_{k}$ the equations of motion for the function $\psi$ analytical
in upper half plane can be linearized by the same substitution.

We suppose to consider more general solutions of eqs \eqref{em} as well as their relations with semiclassical limit of matrix elements of the fields $\Phi_{j}$ in the separate publication.
\section{Concluding remarks}
In this paper we have shown that for spherical geometry the AGT conjecture is valid for the $N=2$ supersymmetric quiver  theories with gauge group 
$U(n)$. To prove this conjecture we considered the algebra $A=W_{n}\otimes H$, which is a tensor product of commuting $W_{n}$ and Heisenberg algebras. We constructed the commuting integrals of motion in the space $A$ which have simple eigenvectors an eigenvalues. The remarkable property of the orthogonal basis of eigenvectors is that all matrix elements of vertex operators in this basis have the factorized form and determine the combinatorial expansion for multipoint conformal blocks. 

It is natural to think that the similar construction can give the combinatorial expansion for all CFT where conformal blocks are completely
defined by the symmetry algebra together with three point functions of primary fields. For example, in $N=1$ and $N=2$ superconformal Liouville theories, parafermionic Liouville CFT (see for example \cite{Bershtein:2010wz}), sine-Liouville model and closely related with it $sl(2)$ WZN model. An evidence of this statement is confirmed by the results of the paper \cite{Awata:2011fk}. 

We note that corresponding integrable systems can have the own interest. They can be related with integrable many body systems. For example, the
integral of motion constructed in this paper describes $n$ Calogero-Sutherland systems coupled in the special way which preserve the
integrability. 

As the simplest example of integrable system in the extended space $A$ we consider $A=G\ast \Psi $, where $G$ is $N=1$ supersymmetry algebra generated by super-current $G(z)$ and stress energy tensor $T(z)$ with the standard commutation relations and $\Psi $ is the algebra generated by the free Majorana fermion $\psi (z)$. In the space $A$ we have two integrals at the level $2$.
\begin{eqnarray}
I_{2} &=&L_{0}+l_{0}  \nonumber \\
\hat{I}_{2} &=&2iQ\sum_{\dot{k}>0}^{\infty }k\psi _{-k}\psi
_{k}+\sum_{-\infty }^{\infty }G_{-k}\psi _{k}  \label{20}
\end{eqnarray}%
here $L_{n}$ and $l_{n}$ are the Fourier components of stress energy tensor $T$ and stress energy tensor of free fermion theory respectively, 
$\psi _{k}$ and $G_{k}$ with half-integer (integer) \ are the modes of the of fields $\psi $ and $G$ in the NS (R) sectors. 
The spectrum of $I_{2}$ is trivial and  $\hat{I}_{2}$ has spectrum linear in momentum $P.$

To show that this system is integrable we give here one next integral
commuting with integrals  $I_{2},\hat{I}_{2},$ which has a form:
\begin{eqnarray}
\hat{I}_{4} &=&\sum_{n+k+l=0}:L_{n}G_{k}:\psi
_{l}+iQ\sum_{k>0}kG_{-k}G_{k}+4iQ\sum_{-\infty }^{\infty }L_{-n}l_{n}+ 
\nonumber \\
&&\frac{(1-14Q^{2})}{4}\sum_{-\infty }^{\infty }k^{2}G_{-k}\psi
_{k}+8iQ^{3}\sum_{\dot{k}>0}^{\infty }k^{3}\psi _{-k}\psi _{k}+\frac{%
iQ(1+2Q^{2})}{8}\sum_{\dot{k}>0}^{\infty }k\psi _{-k}\psi _{k}.  \label{21}
\end{eqnarray}
We think that the integrable systems of this type could play r\^ole in supersymmetric version of AGT relation proposed recently in \cite{Belavin:2011pp} and discussed in  \cite{Belavin:2011tb,Bonelli:2011jx,Bonelli:2011kv}. We propose to discuss this question in a future publication.
\acknowledgments
We thank Sylvain Ribault for discussions and interest to this work. This research was held within the framework of the Federal programs ``Scientific and Scientific-Pedagogical Personnel of Innovational Russia'' on 2009-2013 (state contracts No. P1339 and No. 02.740.11.5165) and was supported by cooperative CNRS-RFBR  grant PICS-09-02-93064  and by Russian Ministry of Science and Technology under the Scientific Schools grant 6501.2010.2. The research of A.L. was also supported in part by the National Science Foundation under Grant No. NSF PHY05-51164 and by Dynasty foundation. 
\appendix
\section{Calculation of the Integral \protect\eqref{sin}}\label{Selberg-integral}
Here we calculate the ratio \eqref{sin}. In our case 
$$
\mathcal{O=}\mathrm{J}_{\mu}^{(1/g)}[p_{k}(t^{(1)})+\eta]\mathrm{J}_{\lambda}%
^{(1/g)}[p_{-k}(t^{(n-1)})].
$$
Following ref \cite{Alba:2010qc} we change the normalization tof Jack polynomials i.e. introduce the polynomials
\begin{equation}
\mathrm{P}_{\lambda}^{(1/g)}(t_{1},\dots ,t_{k})=\frac{1}{\mathrm{c}_{\lambda
}(g)}\mathrm{J}_{\lambda}^{(1/g)}(t_{1},\dots ,t_{k}) \label{P}%
\end{equation}
with
\[
\mathrm{c}_{\lambda}(g)=\prod\limits_{s\in\lambda}
(1+l_{\lambda}(s)+g^{-1}a_{\lambda}(s))
\]
where $a_{\lambda}(s)$ and $\ l_{\lambda}(s)$ are the arm length and the leg
length of the square $s$ in the partition $\lambda$

The polynomials $\mathrm{P}_{\lambda}^{(1/g)}(t_{1},\dots ,t_{k})$ satisfy the
Kadell identity \cite{kadell} which we will use further
\begin{equation}
\mathrm{P}_{\lambda}^{(1/g)}[p_{-m}]=\prod\limits_{j=1}^{k}
t_{j}^{-\lambda_{1}}\mathrm{P}_{\hat{\lambda}}^{(1/g)}[p_{m}] \label{K}%
\end{equation}
where for given partition $\lambda=\left\{  \lambda_{1}\geq\lambda
_{2},\dots \right\}  $ the hatted partition $\hat{\lambda}=\left\{  \hat{\lambda
}_{1}\geq\hat{\lambda}_{2},\dots \right\}  $ is defined by
\begin{equation}
\hat{\lambda}_{j}=\lambda_{1}-\lambda_{k-j+1} \label{parl}%
\end{equation}
We introduce also the generalized Pochammer symbol
\begin{equation}
\left[  z\right]  _{\lambda}=\prod\limits_{j\geq1}(z+(1-j)g)_{\lambda_{j}}=\prod\limits_{j\geq1}
\frac{\Gamma(z+\lambda_{j}+(1-j)g)}{\Gamma(z+(1-j)g)} \label{Ph}%
\end{equation}
and the new notations, which are usual for calculation of $sl(n+1)$ type Selberg
Integrals. Namely,
\begin{equation}
g\rightarrow\gamma,\quad A_{r}\rightarrow\alpha_{r}-1,\quad r=1,\dots  ,n-1,\quad
A_{n}-\lambda_{1}=\alpha_{n}-1,\quad B=\beta-1 \label{gab}%
\end{equation}

It is more convenient to consider the integral not over the collection of the
contours $\mathcal{C}(n+1)$ shown on fig. \ref{contours} but the integral over the special domain
$C_{\gamma}^{k_{1},\dots  ,k_{n}}[0,1]$ \ which will be described bellow. These two
integral are related by overall factor independent on the single-valued
function $\mathcal{O}$ and cancels in the ratio (\ref{sin}). Namely,
\begin{align}
&
\int\limits_{\mathcal{C(}n+1)}
\mathcal{O}(t^{(1)}|\dots |t^{(n)})G_{n+1}(t^{(1)}|\dots  |t^{(n)})d^{k_{1}}%
t^{(1)}\dots  d^{k_{n}}t^{(n)}\nonumber\\
&  =\mathcal{N}\int\limits_{C_{\gamma}^{k_{1},\dots  ,k_{n}}[0,1]}
\mathcal{O}(t^{(1)}|\dots |t^{(n)})|G_{n+1}(t^{(1)}|\dots  |t^{(n)})|d^{k_{1}}%
t^{(1)}\dots  d^{k_{n}}t^{(n)} \label{cci}%
\end{align}
where $\mathcal{N=}N_{k_{1}}(\beta,\gamma)N_{k_{2}}(-k_{1}\gamma
,\gamma)\dots  N_{k_{n-1}}(-k_{n-2}\gamma,\gamma)$. Here $N_{k}(\beta,\gamma)$ is \cite{Tarasov-Varchenko,Warnaar-An}
\[
N_{k}(\beta,\gamma)=\frac{2ie^{i\pi\beta}\sin(\pi(\beta+(j-1)\gamma)\sin(\pi
j\gamma)}{\sin(\gamma)}%
\]
and $G_{n+1}(t^{(1)}|\dots  |t^{(n)})$ is defined by eq \eqref{gt}.

The domain of integration $C_{g}^{k_{1},\dots  ,k_{n}}[0,1]$ for $sl(n+1)$ Selberg
Integrals is defined in the following way \cite{Tarasov-Varchenko,Warnaar-An}. Let all points
$t_{i}^{(r)}$ are ordered
\begin{equation}
0\leq t_{i}^{(r)}\leq t_{2}^{(r)}\leq\dots \leq t_{k}^{(r)}\leq1,r=1,\dots  ,n.
\label{ord}%
\end{equation}
We introduce $n-1$ non-decreasing maps $M_{r}$
\begin{equation}
M_{r}:\left\{  1,\dots  k_{r}\right\}  \rightarrow\{1,\dots  ,k_{r-1}\},\quad
M_{r}(i)\leq M_{r}(i+1),\quad r=2,\dots  n. \label{map}%
\end{equation}
such that $1\leq M_{r}(i)\leq k_{r-1}-k_{r}+i$ and
\begin{equation}
t_{M_{r}(i)-1}^{(r-1)}\leq t_{i}^{(r)}\leq t_{M_{r}(i)}^{(r-1)},\quad r=2,\dots  n.
\label{mmm}%
\end{equation}
where $t_{0}^{(r-1)}=0$. Given admissible maps $M_{r}(i)$ satisfying
eq \eqref{mmm} we define 
$$
D_{M_{1}\dots  M_{n-1}}^{^{k_{1},\dots  ,k_{n}}}[0,1]\subseteq
D^{^{k_{1},\dots  ,k_{n}}}[0,1],
$$ 
where 
$D^{^{k_{1},\dots  ,k_{n}}}[0,1]$ is a domain
(\ref{ord}). As chain we have
\[
D^{^{k_{1},\dots  ,k_{n}}}[0,1]=\sum_{M_{2},\dots  ,M_{n}}D_{M_{2}\dots  M_{n}}%
^{^{k_{1},\dots  ,k_{n}}}[0,1]
\]
The domain of integration $C_{\gamma}^{k_{1},\dots  ,k_{n}}[0,1]$ is defined as
\[
C_{\gamma}^{^{k_{1},\dots  ,k_{n}}}[0,1]=\sum_{M_{2},\dots  ,M_{n}}F_{M_{2}\dots  M_{n}%
}^{k_{1},\dots  ,k_{n}}(\gamma)D_{M_{2}\dots  M_{n}}^{^{k_{1},\dots  ,k_{n}}}[0,1]
\]
where coefficients $F_{M_{2}\dots  M_{n}}^{k_{1},\dots  ,k_{n}}(\gamma)$
\[
\prod\limits_{r=2}^{n}\prod\limits_{i=1}^{k_{r}}
\frac{\sin(\pi(k_{r-1}-k_{r}-M\left(  i)+i+1\right)  \gamma)}{\sin(\left(
\pi(k_{r-1}-k_{r}+i\right)  \gamma)}
\]
coincide up to overall factor $\mathcal{N}$ with the factors which appear in
contour integral in l.h.s. of eq \eqref{cci} (if we tend all contours to the
real axis) in the corresponding domain of integration due to the avoiding of
branching points of the integrand.

To make our transformations more explicit we consider the $sl(3)$ Selberg
integral. The calculation in the case of general $n$ follows the same steps.
In the $sl(3)$ we have only one non-decreasing map $M_{2}=M:\left\{
1,\dots  k_{2}\right\}  \rightarrow\{1,\dots  ,k_{1}\}$ such that $t_{M(i)-1}^{(1)}\leq
t_{i}^{(2)}\leq t_{M(i)}^{(1)}$. The $sl(3)$ Selberg integral has a form
\begin{multline}
\mathbf{J}_{\mu,\lambda}^{(k_{1},k_{2})}(\alpha_{1},\alpha_{2},\beta,\gamma)=\\=
\int\limits_{C_{\gamma}^{^{k_{1},k_{2}}}[0,1]}P_{\mu}^{(1/\gamma)}[p_{m}(t^{(1)})+(\beta-\gamma)/\gamma]
P_{\lambda}^{(1/\gamma)}[p_{m}(t^{(2)})]D^{2\gamma}(t^{(1)})D^{2\gamma}(t^{(2)})\\
|D(t^{(1)},t^{(2)}|^{-\gamma}\prod\limits_{i=1}^{k_{1}}
(t_{i}^{(1)})^{\alpha_{1-1}}(1-t_{i}^{(1)})^{\beta-1}\prod\limits_{j=1}^{k_{2}}
(t_{j}^{(2)})^{\alpha_{1-1}}d^{(k_{1})}t^{(1)}d^{(k_{2})}t^{(2)} \label{ss3}%
\end{multline}
It follows from eqs \eqref{K} and \eqref{gab} that
\begin{equation}
\frac{\left\langle \mathrm{J}_{\mu}^{(1/g)}[p_{k}+\rho]\mathrm{J}_{\lambda
}^{(1/g)}[p_{-k}]\right\rangle _{Sel}^{sl(n)}}{\left\langle 1\right\rangle
_{Sel}^{sl(n)}}=\mathrm{c}_{\lambda}(g)\mathrm{c}_{\mu}(g)\frac{\mathbf{J}%
_{\mu,\hat{\lambda}}^{(k_{1},k_{2})}(1+A_{1},1+A_{2}-\lambda_{1}%
,1+B,g)}{\mathbf{J}_{\scriptscriptstyle{\varnothing},\scriptscriptstyle{\varnothing}}^{(k_{1},k_{2})}(1+A_{1}%
,1+A_{2}-\lambda_{1},1+B,g)} \label{e}%
\end{equation}
where the partition $\hat{\lambda}$ is defined by eq \eqref{parl}.

The calculation of the integral (\ref{ss3}) follows almost exactly the same
steps that in the $sl(2)$ case \cite{Alba:2010qc}. It is based on two identities (A.1
and A.2 in (\cite{Alba:2010qc}) which we reproduce here for completeness. Namely \cite{Okounkov:fk} identity A.1: 
\paragraph{Identity A.1.} Let $\tau^{(2)}=(\tau_{1}^{(2)},\dots ,\tau_{k_{2-1}}^{(2)})$ and
$(t_{1}^{(2)},\dots .,t_{k_{2}}^{(2)})$ satisfy the interlacing property
\[
t_{1}^{(2)}<\tau_{1}^{(2)}<t_{2}^{(2)}<\tau_{2}^{(2)}<\dots .<t_{k_{2}-1}%
^{(2)}<\tau_{k_{2}-1}^{(2)}\,<t_{k_{2}}^{(2)}%
\]
denoted by $\tau^{(2)}\prec t^{(2)}$. Then for $\nu=(\nu_{1}\geq\nu_{2}\dots )$ a
partition of length at most $k_{2}-1$
\begin{equation}
D(t^{(2)})^{2\gamma-1}P_{\nu}^{(1/\gamma)}(t^{(2)})=\Lambda_{\nu}(\gamma)\int\limits_{\tau^{(2)}\prec t^{(2)}}
P_{\nu}^{(1/\gamma)}(\tau^{(2)})D(\tau^{(2)})|D(\tau^{(2)},t^{(2)}%
)|^{\gamma-1}d^{(k_{2}-1)}\tau^{(2)} \label{A1}%
\end{equation}
where
\[
\Lambda_{\nu}(\gamma)=\frac{\Gamma(k_{2}\gamma)}{\Gamma(\gamma)^{k_{2}}}%
\frac{\left[  k_{2}\gamma\right]  }{\left[  (k_{2}-1)\gamma\right]  }%
\]
The second identity states:
\paragraph{Identity A.2.} Let \ $t^{(1)}=($\ $t_{1}^{(1)}\dots t_{k_{1}}^{(1)})$ and $\tau^{(1)}%
=(\tau_{1}^{(1)},\dots ,\tau_{k_{1-1}}^{(1)})$ satisfy the interlacing property
\[
0<t_{1}^{(1)}<\tau_{1}^{(1)}<t_{2}^{(1)}<\tau_{2}^{(1)}<\dots .<t_{k_{1}-1}%
^{(1)}<\tau_{k_{1}-1}^{(1)}\,<t_{k1}^{(1)}<1
\]
denoted by $t^{(1)}\prec\tau^{(1)}$. Then for partition $\mu=(\mu_{1}\geq
\mu_{2}\dots )$ a partition
\begin{align}
&\int\limits_{t^{\left(  1\right)  }\prec\tau^{\left(  1\right)
}}
P_{\mu}^{(1/\gamma)}[p_{m}(t^{(1)})+(\beta-\gamma)/\gamma]D(t^{(1)}%
)|D(t^{(1)},\tau^{(1)})|^{\gamma-1}\prod\limits_{i=1}^{k_{1}}
(t_{i}^{(1)})^{\alpha_{1-1}}(1-t_{i}^{(1)})^{\beta-1}d^{k_{1}}t^{(1)}
\nonumber\\
&  =\Xi_{\mu}(\alpha_{1},\beta,\gamma)\prod\limits_{i=1}^{k_{1}-1}
(\tau_{i}^{(1)})^{\alpha_{1-1}}(1-\tau_{i}^{(1)})^{\beta-1}D(\tau
^{1})^{2\gamma-1}P_{\mu}^{(1/\gamma)}[p_{m}(\tau^{(1)})+\beta/\gamma]
\label{A2}%
\end{align}
where
\[
\Xi_{\mu}(\alpha_{1},\beta,\gamma)=\frac{\Gamma(\alpha_{1})\Gamma(\beta
)\Gamma^{k_{1}-1}(\gamma)}{\Gamma(\alpha_{1}+\beta+(k_{1}-1)\gamma)}%
\frac{(\alpha_{1}+\beta+(k_{1}-2)\gamma)_{\mu}}{(\alpha_{1}+\beta
+(k_{1}-1)\gamma)_{\mu}}%
\]
For our purposes we need also third identity which is the analog of identity
derived in \cite{Baseilhac:1998eq} for the complex integrals (i.e. for integrals over all plane
for each of variables). For our case this identity can be formulated as
\begin{multline}
\int\limits_{\tau^{(2)}\prec t^{(2)}}\int\limits_{C_{\gamma}^{^{k_{1},k_{2}}}[0,1]}\prod\limits_{i=1}^{k_{2}}
(t_{i}^{(2)})^{\alpha_{2}-1}D(t^{(2)})|D(\tau^{(2)},t^{(2)})|^{\gamma-1}|D(t^{(1)},t^{(2)})|^{-\gamma}d^{k_{2}}t^{(2)}=\\=
\Omega_{k_{1},k_{2}}(\alpha_{2},\gamma)D(\tau^{(2)})^{2\gamma-1}D(t^{(1)})^{1-2\gamma}\prod\limits_{i=1}^{k_{2}-1}
(\tau_{i}^{(2)})^{\alpha_{2}+\gamma-1}\prod\limits_{j=1}^{k_{1}}
(t_{j}^{(1)})^{\alpha_{2}-\gamma}\\
\int\limits_{t^{\left(  1\right)  }\prec\tau^{\left(  1\right)
}}\int\limits_{C_{\gamma}^{^{k_{1}-1,k_{2}-1}}[0,1]}
\prod\limits_{i=1}^{k_{1}-1}
(\tau_{i}^{(1)})^{-\alpha_{2}}D(\tau^{(1)})|D(\tau^{(1)},\tau^{(2)})|^{-\gamma}|D(t^{(1)},\tau^{(1)})|^{\gamma
-1}d^{(k_{1}-1)}\tau^{(1)} \label{A3}%
\end{multline}
where
\[
\Omega_{k_{1},k_{2}}(\alpha_{2},\gamma)=\frac{\pi\Gamma(\alpha_{2}%
)\Gamma^{k_{2}-k_{1}-1}(\gamma)}{\sin(\pi k_{1}\gamma)\Gamma(1+\alpha
_{2}-(k_{2}-k_{1}+1)\gamma)}%
\]
Now we can reduce the number of integrations in \eqref{ss3} performing the following steps
\begin{enumerate}
\item For $\lambda=(\lambda_{1},\dots  ,\lambda_{k_{2}})$ we use the relation
\[
P_{\nu}^{(1/\gamma)}(t^{(2)})=P_{\lambda}^{(1/\gamma)}[p_{m}(t^{(2)}%
)]=(t_{1}^{(2)}\dots t_{k_{2}}^{(2)})^{\lambda_{k_{2}}}P_{\nu}^{(1/\gamma)}%
[p_{m}(t^{(2)})]
\]
where $\nu=(\lambda_{1}-\lambda_{k_{2}},\dots ,\lambda_{k_{2}-1}-\lambda_{k_{2}%
},0)$
\item We represent $D(t^{(2)})^{2\gamma-1}P_{\nu}^{(1/\gamma)}(t^{(2)})$ in
eq.(\ref{ss3}) using (\ref{A1})
\item We use the integral relation (\ref{A3})
\item We compute the integral over variables $t^{(1)}$ using (\ref{A2}).
\end{enumerate}

As a result we reduce our integral (\ref{ss3}) to the integral of the same
form but with lower number of integrations $(k_{1},k_{2})\rightarrow
(k_{1}-1,k_{2}-1)$
\begin{multline}
\mathbf{J}_{\mu,\lambda}^{(k_{1},k_{2})}(\alpha_{1},\alpha_{2},\beta,\gamma)=\\=
\frac{\left[  k_{2}\gamma\right]  \Gamma(k_{2}\gamma)\Gamma(\alpha_{2}+\lambda_{k_{2}})\Gamma(\hat{\alpha}_{12}+\lambda_{k_{2}}-\gamma
)\Gamma\left(  \beta\right)  \Gamma(1-k_{1}\gamma)\Gamma(k_{1}\gamma)}{\left[
(k_{2}-1)\gamma\right]  \Gamma(\gamma)^{2}\Gamma(1+\alpha_{2}+\lambda_{k_{2}%
}-(k_{1}-k_{2}+1)\gamma)\Gamma(\hat{\alpha}_{12}+\beta+\lambda_{k_{2}}%
+(k_{1}-2)\gamma)}\\
\frac{[\hat{\alpha}_{12}+\beta+\lambda_{k_{2}}+(k_{!}-3)\gamma]_{\mu}%
}{[\hat{\alpha}_{12}+\beta+\lambda_{k_{2}}+(k_{1}-2)\gamma]_{\mu}}\,
\mathbf{J}_{\mu,\nu}^{(k_{1}-1,k_{2}-1)}(\alpha_{1},\alpha_{2}+\gamma
,\beta+\gamma,\gamma) \label{ikr}%
\end{multline}
where $\nu=(\lambda_{1}-\lambda_{k_{2}},\dots \lambda_{k_{2}-1}-\lambda_{k_{2}%
},0)$. Now we can proceed by induction, We should be careful, because after
$k_{2}$ steps our integral transforms to the $sl(2)$ Selberg integral
$\mathbf{J}_{\mu}^{(k_{1}-k_{2})}(\alpha_{1},\beta+k_{2}\gamma,\gamma)$ and
further steps of induction give
\[
\mathbf{J}_{\mu}^{(k)}(\alpha,\beta^{\prime},\gamma)=\frac{\Gamma
(\alpha)\Gamma\left(  \beta^{\prime}\right)  \Gamma(k\gamma)[\alpha
+\beta+(k-2)\gamma]_{\mu}}{\Gamma(\alpha+\beta^{\prime}+(k-1)\gamma
)\Gamma(\gamma)[\alpha+\beta+(k-1)\gamma]_{\mu}}\mathbf{J}_{\mu}%
^{(k-1)}(\alpha+\gamma,\beta^{\prime}+\gamma,\gamma)
\]
Using that
\[
\frac{\left[  k_{2}\gamma\right]  _{\nu}}{\left[  (k_{2}-1)\gamma\right]
_{\nu}}=\frac{P_{\lambda}^{(1/\gamma)}\left[  k_{2}\right]  }{P_{\nu
}^{(1/\gamma)}\left[  k_{2}-1\right]  }%
\]
we find
\begin{multline}\label{imlk}
\mathbf{J}_{\mu,\lambda}^{(k_{1},k_{2})}   =P_{\lambda}^{(1/\gamma)}\left[
k_{2}\right]  P_{\mu}^{(1/\gamma)}\left[  k_{1}+(\beta-\gamma)/\gamma\right]
\prod\limits_{j=1}^{k_{1}}
\frac{\Gamma(\beta+\left(  j-1\right)  \gamma)\Gamma\left(  j\gamma\right)
}{\Gamma\left(  \gamma\right)}\\
\prod\limits_{j=1}^{k_{1}-k_{2}}
\frac{\Gamma(\alpha_{1}+\left(  j-1\right)  \gamma)}{\Gamma(\alpha_{1}
+\beta+(k_{1}+j-2)\gamma)}
\frac{\lbrack\alpha_{1}+\beta+(k_{1}+j-3)\gamma]_{\mu}}{[\alpha_{1}%
+\beta+(k_{1}+j-2)\gamma]_{\mu}}\\
\prod\limits_{j=1}^{k_{2}}
\frac{\Gamma(\alpha_{2}+\left(  k_{2}-j\right)  \gamma+\lambda_{j})}%
{\Gamma(1+\alpha_{2}+(2k_{2}-k_{1}-j-1)\gamma+\lambda_{j})\Gamma\left(
\gamma\right)}\\
\frac{\Gamma(\hat{\alpha}_{12}+(k_{2}-j-1)\gamma+\lambda_{j})\Gamma\left(
j\gamma\right)  \Gamma\left(  1-\left(  k_{1}-k_{2}+j\right)  \gamma\right)
}{\Gamma(\hat{\alpha}_{12}+\beta+(k_{1}+k_{2}-j-2)\gamma+\lambda_{j})}\\
\frac{(\hat{\alpha}_{12}+\beta+(k_{1}+k_{2}-j-3)\gamma+\lambda_{j})_{\mu}%
}{(\hat{\alpha}_{12}+\beta+(k_{1}+k_{2}-j-2)\gamma+\lambda_{j})_{\mu}}
\end{multline}
here $\hat{\alpha}_{12}=\alpha_{1}+\alpha_{2}$.

Now we can calculate the ratio (\ref{e}). Using evaluation formula \cite{Stanley}
\[
P_{\lambda}^{(1/\gamma)}\left[  N\right]  =\gamma^{-|\lambda|}\frac{\left[
N\gamma\right]  }{\mathrm{c}_{\lambda}(g)}%
\]
and eqs \eqref{e}, \eqref{parl} and \eqref{imlk} we find that the ratio $\frac{\left\langle
\mathrm{J}_{\mu}^{(1/g)}[p_{k}+\rho]\mathrm{J}_{\lambda}^{(1/g)}%
[p_{-k}]\right\rangle _{Sel}^{sl(3)}}{\left\langle 1\right\rangle
_{Sel}^{sl(3)}}$ is
\begin{multline}
g^{-|\lambda|-|\mu|}\left[  k_{2}g\right]  _{\lambda}\left[  1+B+(k_{1}%
-1)g\right]  _{\mu}\\
\prod\limits_{j=1}^{k_{2}}
\frac{\Gamma(1+A_{2}+\left(  j-1\right)  g-\lambda_{j})\Gamma(2+\hat{A}%
_{12}+(j-2)g-\lambda_{j})}{\Gamma(1+A_{2}+\left(  j-1\right)  g)\Gamma
(2+\hat{A}_{12}+(j-2)g)}\\
\frac{\Gamma(2+A_{2}+(k_{2}-k_{1}+j-2)g)\Gamma(3+\hat{A}_{12}%
+B+(k_{1}+j-3)g)}{\Gamma(2+A_{2}+(k_{2}-k_{1}+j-2)g-\lambda_{j})\Gamma
(3+\hat{A}_{12}+B+(k_{1}+j-3)g-\lambda_{j})}\\
\frac{[3+\hat{A}_{12}+B+(k_{1}+j-4)g-\lambda_{j}]_{\mu}}{[3+\hat{A}%
_{12}+B+(k_{1}+j-3)g-\lambda_{j}]_{\mu}}\frac{[2+A_{1}+B+(k_{1}-2)g]_{\mu}%
}{[2+A_{1}+B+(2k_{1}-k_{2}-2)g]_{\mu}} \label{otv}%
\end{multline}
After simple transformations the r.h.s. of this equality can be rewritten as
\begin{multline}
\hspace*{-8pt}\frac{g^{-|\lambda|-|\mu|}\left[  1+B+(k_{1}-1)g\right]  _{\mu}%
[2+A_{1}+B+(k_{1}-2)g]_{\mu}\left[  k_{2}g\right]  _{\lambda}\left[
-1-A_{2}+(k_{1}-k_{2}+1)g\right]  _{\lambda}}{[2+A_{1}+B+(2k_{1}%
-k_{2}-2)g]_{\mu}[3+\hat{A}_{12}+B+(k_{1}+k_{2}-3)g]_{\mu}\left[
-A_{2}\right]  _{\lambda}\left[  -1-\hat{A}_{12}+g\right]  _{\lambda}}\\
\prod\limits_{s\in\lambda}
(2+\hat{A}_{12}+B+(k_{1}-3)g-a_{\mu}(s)g-l_{\lambda}(s))\\
\prod\limits_{t\in\mu}(3+\hat{A}_{12}+B+(k_{1}-2)g+a_{\lambda}(t)g+l_{\mu}(s)) \label{res}%
\end{multline}
We note that screening condition (\ref{sc}) can be written in $sl(3)$ case as
\[
-x_{1}+x_{1}^{\prime}+2a/3+k_{1}b=0,\quad -x_{2}+x_{2}^{\prime}-a/3+(k_{2}-k_{1})b=0,\quad -x_{3}+x_{3}^{\prime}-a/3-k_{2}b=0.
\]
If we take any two from these three equations to express $k_{1},k_{2}$ in
terms of $x,x^{\prime}$ and $a$ and take into account eq \eqref{AB} we derive
desirable result \eqref{sin}.

The calculation of $sl(n+1)$ -Selberg integral $\mathbf{J}_{\mu,\lambda
}^{(k_{1},\dots  ,k_{n})}(\vec{\alpha},\beta,\gamma)$ (here $\vec{\alpha}%
=(\alpha_{1},\dots  ,\alpha_{n})$) follows exactly the same steps. We should only
repeat the integral relation (\ref{A3}) $n-1$ times. In this way we derive the relation
\begin{multline*}
\mathbf{J}_{\mu,\lambda}^{(k_{1},\dots  ,k_{n})}(\vec{\alpha},\beta,\gamma) =\\=
\frac{\left[  k_{n}\gamma\right]  \Gamma(k_{2}\gamma)\Gamma(\alpha
_{n}+\lambda_{k_{n}})\Gamma\left(  \beta\right)  [\hat{\alpha}_{1n}%
+\beta+(k_{1}-n-1)\gamma+\lambda_{k_{n}}]_{\mu}}{\left[  (k_{n}-1)\gamma
\right]  \Gamma(\gamma)^{n}\Gamma\left(  \hat{\alpha}_{1n}+\beta
+(k_{1}-n\right)  \gamma+\lambda_{k_{n}})[\hat{\alpha}_{1n}+\beta
+(k_{1}-n\gamma+\lambda_{k_{n}}]_{\mu}}\\
\prod\limits_{j=1}^{n-1}
\frac{\Gamma(1-k_{j}\gamma)\Gamma(k_{j}\gamma)\Gamma\left(  \hat{\alpha
}_{n-j,n}-j\gamma+\lambda_{k_{n}}\right)  }{\Gamma\left(  1+\hat{\alpha
}_{n-j+1,n}-(k_{n-j}-k_{n-j+1}+1)\gamma+\lambda_{k_{n}}\right)  }%
\mathbf{J}_{\mu,\nu}^{(k_{1}-1,\dots  ,k_{n}-1)}(\vec{\alpha}^{\prime},\beta
+\gamma,\gamma)
\end{multline*}
where $\nu=(\lambda_{1}-\lambda_{k_{n}},\dots ,\lambda_{k_{n}-1}-\lambda_{k_{n}},0)$, $\hat{\alpha}_{r,n}=\alpha_{r}+\dots +\alpha_{n}$, $\hat{\alpha}_{n,n}=\alpha_{n}$ and $\vec{\alpha}^{\prime}=(\alpha_{1},\dots ,\alpha_{n-1},\alpha_{n}+\gamma)$.

If we solve this recurrence relation and use the screening conditions to express $k_{1},\dots ,k_{n}$ in terms of terms of $x,x^{\prime}$ and $a$ we derive the expected result for matrix element \eqref{sin}.
\bibliographystyle{MyStyle} 
\bibliography{MyBib}
\end{document}